\documentclass[aps, prd, groupedaddress, eqsecnum, nofootinbib, showpacs, preprintnumbers]{revtex4}
\usepackage{graphicx, epsf, amssymb, amsbsy, amsfonts, amssymb, amsmath}
\usepackage[usenames, dvipsnames]{color}

\usepackage{calc}
\usepackage{url}

\newif\ifdraft
\drafttrue
\newif\ifpreprint
\preprinttrue

\newif\iffigs\figstrue

\DeclareFontFamily{U}{rsf}{}
\DeclareFontShape{U}{rsf}{m}{n}{
  <5> <6> rsfs5 <7> <8> <9> rsfs7 <10-> rsfs10}{}
\DeclareMathAlphabet\Scr{U}{rsf}{m}{n}

\def\tab#1{table~{\ref{#1}}}

\def\sect#1{section~{\ref{#1}}}

\def\app#1{appendix~{\ref{#1}}}
\def\eqn#1{eq.~(\ref{#1})}

\newcommand{\bea}{\begin{eqnarray}}
\newcommand{\eea}{\end{eqnarray}}
\newcommand{\be}{\begin{equation}}
\newcommand{\ee}{\end{equation}}
\newcommand{\non}{\nonumber}

\def\cN{{\cal N}}

\def\cD{{\cal D}}

\def\cS{{\cal S}}

\def\cL{{\Scr L}}

\def\a{\alpha}

\def\b{\beta}

\def\d{\delta}

\def\g{\gamma}

\def\k{\kappa}

\def\o{\omega}

\def\q{\theta}

\newcommand{\ad}{{\dot{\alpha}}}                           
  
\newcommand {\cO}{{\cal O}}                          

\newcommand{\pa}{\partial}                           
\newcommand{\hf}{\frac12}

\newcommand {\cW}{{\cal W}}

\newcommand {\cM}{{\cal M}}

\newcommand {\cZ}{{\cal Z}}

\newcommand{\cH}[1]{ {{\cal H}{}^{(#1)}}}
\newcommand{\cHb}[1]{ { { \overline{\cal H } }{}^{(#1)}}  }

\newcommand {\cZbar}{ { \overline {\cal Z} } }

\newcommand {\cMbar}{ {\overline {\cal M} }}

\newcommand {\cWbar}{ {\overline {\cal W} }}

\newcommand {\SD}{ { {\cal D}^4 } }

\newcommand {\SDbar}{ { {\overline {\cal D}}^4 } }

\def\NeqEight{{\cal N}=8}
\def\NeqTwo{{\cal N}=2}
\def\NeqOne{{\cal N}=1}

\begin{document}

\preprint{SU-ITP-12/05}

\preprint{CERN-PH-TH/2012-006}

\title{  $\boldsymbol{{\cal N}=2}$ Supersymmetry and $\boldsymbol{U(1)}$-Duality }

\

\

\author{\bf Johannes Broedel${}^a$, John Joseph M. Carrasco${}^a$, Sergio Ferrara${}^b$, Renata Kallosh${}^a$, and Radu Roiban${}^c$}

\affiliation{\vskip .43cm
${}^a$Stanford Institute for Theoretical Physics and Department of Physics, Stanford University, 
Stanford, CA 94305-4060, USA\\
${}^b$Physics Department, Theory Unit, CERN, 1211 Geneva 23, Switzerland and
INFN, Laboratori Nazionali di Frascati, Via Enrico Fermi, 40,  00044 Frascati, Italy and \\
Department of Physics and Astronomy, University of California, Los Angeles, CA 90095-1547,USA\\
${}^c$Department of Physics, Pennsylvania State University, University Park, PA 16802, USA
}

\begin{abstract}

Understanding the consequences of the $E_{7(7)}$ duality on the UV properties of $\NeqEight$ supergravity requires unravelling when and how duality-covariant actions can be constructed so as to accommodate duality-invariant counter-terms.
 For non-supersymmetric abelian gauge theories exhibiting $U(1)$-duality, with and without derivative couplings, it was shown that such a covariant construction is always possible. In this paper we describe a similar procedure for the construction of covariant non-linear deformations of $U(1)$-duality invariant theories  in the presence of rigid ${\cal N}=2$ supersymmetry.
This is a concrete step towards studying the interplay of duality and extended supersymmetry.

\end{abstract}

\pacs{04.65.+e, 11.15.Bt, 11.30.Pb, 03.50.-z, 03.50.De \hspace{1cm}}

\maketitle


\newpage
\setcounter{page}{0}
\tableofcontents
\thispagestyle{empty}
\clearpage

\pagestyle{plain}
\setcounter{page}{1}
\section{Introduction}
When the  equations of motion of a classical field theory are said to respect a
duality invariance, they are invariant under the rotation of an electric field
into its magnetic field or of  a field strength into its dual field strength.  
Among the theories exhibiting such properties are Maxwell's theory, whose
duality group is $U(1)$, and extended  ${\cal N}\geq 2$ supergravity theories,
whose duality properties were first discussed in \cite{Ferrara:1976iq}.
Of recent interest is  the maximally supersymmetric supergravity
theory~\cite{Cremmer:1978km}, ${\cal N}=8$ supergravity, whose duality group is
$E_{7(7)}$.  Such dualities could have rather nontrivial consequences.  Indeed,
it has been suggested that the UV-finiteness  of ${\cal N}=8$ supergravity
through four loops \cite{N8Calculation} could be explained by the absence of
possible $E_{7(7)}$-invariant counterterms \cite{Brodel:2009hu,Beisert:2010jx}.

The requirement that  duality invariance be respected is very constraining,
as it relates free and interaction terms.  Perturbatively, the duality symmetry
is expected to be continuous; as such, it has an associated conserved current,
known as the Noether-Guillard-Zumino (NGZ) current \cite{Gaillard:1981rj}.  The
conservation of the NGZ current leads to non-trivial constraints on the possible
deformations of the theories, as the addition of duality-invariant terms does
not generically preserve the duality invariance of the equations of motion.  An
essential ingredient in these constraints is the fact that the dual field
strengths are determined by the equations of motion which receive contributions
from the deformation terms and thus modify the NGZ current.  In ${\cal N}=8$
supergravity, the relation between $E_{7(7)}$ invariants and the conservation of
the duality current was  raised in \cite{Kallosh:2011dp}.

A very natural and interesting question is whether and in what sense classical
duality symmetries are preserved at  the quantum level. A direct answer to this
question is not  straightforward; for example, it is not clear how duality
symmetries are generically visible in scattering amplitudes.   While  multi-soft
scalar limits  can probe \cite{Bianchi:2008pu} the coset structure of
supergravity theories, it is not immediately clear how to similarly probe 
the transformation properties of vector fields.
Several indirect approaches are  possible.  One could construct the (local part
of the) effective action and see if the equations of motion derived from it obey
the same duality symmetry as at the classical level.  Another approach would be
to construct rational functions of momenta obeying the properties of scattering
amplitudes which also obey the multi-soft scalar limit constraints. Both
approaches have been explored in \cite{Kallosh:2011dp} and
\cite{Brodel:2009hu,Beisert:2010jx}, respectively, for constraining and
constructing possible  counterterms of $\NeqEight$ supergravity. 

There exist examples of duality transformations which receive modifications at
the quantum level. For two-dimensional sigma models, T-duality parallels
electric/magnetic duality of four dimensional gauge gauge theories. In addition
to replacing a field by its dual, invariance under duality transformation also
requires changes of the parameters of the theory ({\it i.e.} of the target space
supergravity fields). It was shown in  \cite{Tseytlin:1991wr} that the T-duality
transformation rules \cite{Buscher:1987qj} which guarantee the duality
invariance of sigma models at one-loop level should be modified at higher loops.
These higher-loop corrections may also be reinterpreted as higher-loop
corrections to the relation between the sigma model fields and their duals.
It is thus important to keep in mind the possibility of similar corrections in
more general duality-satisfying interacting quantum field theories.

In order for a  theory to preserve, at the quantum level, the duality of its
classical equations of motion, while admitting a duality-invariant counterterm,
it is necessary that it admits higher-order deformations that maintain the
action's duality covariance.  Recently, Bossard and Nicolai suggested~\cite{BN}
that there exist algorithms to perturbatively deform all  duality-satisfying
theories in a manner consistent with the classical duality transformations.  If
true, besides offering a possibility of constraining the finiteness of ${\cal
N}=8$ supergravity~\cite{Kallosh:2011dp, BN, CKR}, it suggests the possibility
of constructing non-trivial Born-Infeld-type supergravity theories the first of
which would be ${\cal N}=2$ supergravity, as proposed in~\cite{CKR}.  

The procedures outlined in~\cite{BN} involved adding one nonlinear initial
deformation source to the consistency relations imposed by the tree-level
duality transformations and solving them -- resulting in an infinite number of
terms contributing to the effective action.  Such consequences are in line with
expectations based on soft scalar limits \cite{Beisert:2010jx} in the case of
$\NeqEight$ supergravity.  The unmodified covariant  procedure of~\cite{BN},
however, does not reproduce known simpler duality-satisfying effective actions.
In~\cite{CKR} three of the current authors explored the
single-deformation-source approach and found that, in general, an infinite
series of deformations of the consistency relations are required to reproduce
known results.  
Indeed, obtaining even the venerable Born-Infeld (BI) model in the absence of
supersymmetry requires an infinite sequence of deformation sources to the
consistency relations.  A generalized procedure was therefore proposed and
analyzed for bosonic theories with no explicit derivatives.  The algorithms
developed in~\cite{CKR} were used in \cite{CKO} to construct a class of
self-dual models which, in addition to standard BI terms $F^n$, include higher
derivatives terms $\partial ^{4n} F^{2n+2}$.

A necessary step for the extension of this procedure to supergravity theories
(and for showing that it is indeed possible to preserve the classical duality
symmetry in the presence of quantum corrections) is the construction  of
supersymmetric theories of vector multiplets exhibiting duality symmetries. Such
actions have been constructed previously through different methods: a manifestly
supersymmetric Born-Infeld model was constructed in \cite{Cecotti:1986gb},
non-linear superfield actions with spontaneously broken supersymmetry were
studied in \cite{Bagger:1996wp, Ket1, Ket2}, and  models with manifest
supersymmetry and non-linear electromagnetic duality were developed in
\cite{Kuzenko:2000tg, Kuzenko:2000uh, Bellucci:2001hd, Ketov:2001dq}.

In this paper we describe the application of the constructions of~\cite{BN, CKR}
to theories with rigid ${\cal N}=2$ supersymmetry when the duality is of
$U(1)$-type. It is important to stress that this setup is different from the one
of supergravity theories exhibiting duality symmetries. Indeed, here the
starting action is free and the deformation may be tuned as desired; if the
tree-level action is interacting, the deformation is generated by quantum
corrections within this theory and cannot be freely adjusted. While the former
setup is far less constraining than the latter, the construction of deformations
of free actions has proved in the past not to be straightforward.

After setting up a convenient notation, in \sect{Sct:Review} we briefly review
the generalized procedure of \cite{CKR}, and the known  duality-consistent
models in the context of ${{\cal N}=2}$  supersymmetry.  
In \sect{Sct:Action} we describe a new form of the action in terms  of
corrections generated by a deformation source.  This  allows us to write the
action directly in terms of a recursively solved non-linear constraint.   We
provide examples of specific sources in  \sect{Sct:Examples} and discuss how
they can be simply combined to generate a wide variety of actions, including the
BI action found in the literature.   We conclude and comment on the next steps
required towards approaching the construction of a  Born-Infeld-type ${\cal
N}=2$ supergravity and potential ramifications for ${\cal N}=8$ supergravity in
\sect{Sct:Discussion}. In \app{appA}  we provide, for completeness,  information
on ${\cal N}=0$ and ${\cal N}=1$ duality invariant models. In \app{appB} we
argue that under ${\cal N}=2$ supersymmetry, in contrast with ${\cal N}=1$, the
electromagnetic duality models require the presence of space-time derivatives
acting on superfields.  In \app{conventionsSUSY} we provide a summary of our
${\cal N}=2$ superspace conventions.

\section{Review \label{Sct:Review} } 

Let us begin by recalling the covariant construction \cite{CKR} of duality-satisfying actions in terms of 
deformation sources. We will then proceed to summarize the known ${\cal N}=2$ supersymmetric theories 
whose equations of motion are duality-invariant.

\subsection{Generalized duality covariant procedure \label{Sct:Review_CKR}}

To efficiently formulate classical duality relations, and their 
corrections, it is useful to organize fields $F$ and dual fields $G$ such that
the classical duality transformations act as simply as possible. For Maxwell's
theory, as well as supersymmetric versions, this duality transformation can be
expressed as a simple multiplication by a phase $B$ and the relevant complex
field basis is 
\be
T=F-{\rm i}\,G~,~~{\overline T}=F+{\rm i}\,G
\ee
or rather their self-dual and anti-self-dual components $T^\pm =
\frac{1}{2}(T\pm {\rm i}{\widetilde T})$ and ${\overline T}^{\pm} =
\frac{1}{2}({\overline T} \pm {\rm i} {\widetilde {\overline T}})$ as follows
\be
\delta T^\pm={\rm i}\,B \,T^\pm ~~~~\delta {\overline T}{}^\pm=-{\rm i}\,B\, {\overline T}{}^\pm \,.
\label{dualityU1}
\ee
Throughout this paper we suppress  space-time indices whenever possible.
However, introducing the tilde operation, as we do in defining $T^\pm$,
involves normalized contraction with the Levi-Civita symbol, 
\be
\widetilde{A}^{\mu
\nu}\equiv\frac{1}{2}\epsilon^{\mu \nu \rho \sigma} A_{\rho \sigma}\,.
\ee  A similar
organization of fields holds for supergravity theories; the main difference is
the appearance of scalar field-dependent matrices in the analogs of the
expressions above \cite{Gaillard:1981rj, Andrianopoli:1996ve}.
In terms of these fields, the undeformed linear duality constraint on the equations of motion of field strengths 
$T$ can be given quite simply by a ``twisted self-duality'' constraint:
\be
 T^{+}= {\overline T}{}^{-}=0\,.
\label{linsd}
\ee
In these variables, the constraint \cite{Gaillard:1981rj} that the action be self-dual is
\be
{\overline T}{}^{+}T^+-{\overline T}{}^{-}T^-=0 ~ .
\label{NGZbose}
\ee
The twisted self-duality relation~(\ref{linsd}) 
may be interpreted as more fundamental than the action.  Indeed, the action can be determined
by its relation to the dual field strength.  In the case of gauge 
theories this relation is just
\be
G = 2\frac{\delta \cS}{\delta F}
\label{Gdef}
\ee 
For supergravity theories the fields $F$ and $G$ (and thereby $T$) acquire further indices, specifying their 
transformation under the 
duality group, e.g.  $T\mapsto T^{AB}$, {\it etc}. In the following we will not write such indices.

A covariant procedure proposed in \cite{CKR}, generalizing that of \cite{BN}, parametrizes the possible 
deformations of an action in terms of a function ${\cal I}(T^-, {\overline T}^{+},\lambda)$ where $\lambda$ is a dimensionful 
coupling constant.  We start with a duality conserving initial action $\cS_{\rm initial}$, and a duality-invariant  
counterterm (or deformation) $\Delta \cS$,  expressible as a function of the 
conjugate self-dual field-strength $\overline{T}{}^{+}$.  We wish to arrive at an action ${\cal S}_{\rm final}$ 
that incorporates the counterterm yet still conserves the duality current.  We proceed as follows~\cite{CKR}:
\begin{enumerate}

\item Take the variation of the counterterm with respect to the field-strength,  and express as a  function 
of $T^-$, and $\overline{T}{}^+$, 
\be 
\frac{\delta \Delta \cS}{\delta {\overline T}{}^{+}} \to 
\frac{\delta {\cal I}(T^-_{}, {\overline T}{}^{+ },\lambda)}{ \delta {\overline T}{}^{+}}\,.
\ee

\item Introduce an ansatz for the deformation source ${\cal I}( T^{-}_{}, {\overline T}{}^{+},\lambda)$. 
In general, this may be taken to depend on all possible duality invariants.

\item Constrain the self-dual field strength to this variation: 
\be
\label{defGenA}
T^+_{}=\frac{\delta {\cal I}( T^{-}_{}, {\overline T}{}^{+},\lambda)}{ \delta {\overline T}{}^{+}}\,.
\ee

\item Translate \eqn{defGenA} to a differential constraint  on ${\cal S}_{\rm final}$.

\item Introduce an ansatz for ${\cal S}_{\rm final}$, which is analytic around the origin, in terms of 
Lorentz invariants constructed from 
$T^-$ and $\overline{T}{}^+$.  For the case of $U(1)$, as we will see in \sect{Sct:Action}, 
this will be straightforward for ${\cal N}=2$ abelian gauge theories.  In general this is unknown and can depend on 
other fields (e.g. scalars) in non-trivial ways.  

\item Solve for both the ${\cal I}$ ansatz  parameters, as well as the 
final action ansatz parameters, order by order in the coupling constant, enforcing 
the consistency of the relevant NGZ  consistency equation  and any additional desired 
symmetries of the target action, enlarging the ansatz if one runs into inconsistency.
\end{enumerate}

The duality conservation relation (\ref{NGZbose}) imposes constraints on the possible deformation sources. 
If the deformation source ${\cal I}(T, {\overline T},\lambda)$ is hermitian this constraint is simply that ${\cal I}$ be
invariant under the duality transformation, \eqn{dualityU1},
\be
\left({\overline T}^+ \frac{\delta }{ \delta {\overline T}{}^{+}}
-
{T}^- \frac{\delta }{ \delta {T}{}^{-}} \right){\cal I}( T^{-}_{}, {\overline T}{}^{+},\lambda)= 0 \ .
\ee
This differential operator ``measures'' the charge of ${\cal I}$ under the
$U(1)$ duality transformation (\ref{dualityU1}) and thus requires that ${\cal
I}$ is invariant.  As discussed in \cite{CKR}, an invariant deformation source
does not imply that the deformation of the action is also invariant; rather, as
discussed in \cite{Gaillard:1981rj}, the complete deformed action should
transform nontrivially under duality transformations.

In sections~\ref{Sct:Action} and \ref{Sct:Examples} we will see that these steps
have natural counterparts  in models with ${\cal N}=2$  rigid supersymmetry;
they will allow us to easily  superpose  arbitrary $U(1)$-invariant deformation
sources\footnote{Here $U(1)$ invariance refers to invariance under duality
transformations which, in terms of the field variables $T$ and $T^*$, act as
$U(1)$ transformations.} and will lead to a straightforward generation of new
classes of models -- as well as a systematic reconstruction of known ones.

\subsection{${\cal N}=2$ Supersymmetry and Duality\label{Sct:Review_Neq2} }

The interplay between supersymmetry and duality invariance has been studied at length in the literature. A derivation and a review of the main results are given by Kuzenko and Theisen in \cite{Kuzenko:2000uh}.
While it is known how to promote essentially every duality-satisfying bosonic model to an 
${\cal N}=1$ supersymmetric one, 
adding further supercharges  proved to be relatively difficult.  ${\cal N}=2$ supersymmetric extensions of the 
BI theory have been found \cite{Ket1, Ket2,Kuzenko:2000uh, Ketov:2001dq,Bellucci:2001hd}; one of their 
essential features is the presence of  explicit spacetime superfield
derivatives in the action. We will review here  these actions; in appendix~B we will argue that any such 
action necessarily contains spacetime derivatives, in addition to spinorial derivatives of the superfields.

In the absence of hypermultiplets, the standard ${\cal N}=2$ superspace provides an effective framework
for organizing actions and their deformations.  It is parametrized by four bosonic and 
eight fermionic coordinates
$\cZ^A = (x^a, \q^\a_i, {\bar \q}^i_\ad) $, with $a$ and $\a$ being a vector and Weyl spinor Lorentz 
indices and $i = {1}, {2}$  being the $SU(2)$ R-symmetry  index.
Actions describing the dynamics of ${\cal N}=2$ vector multiplets  are written in terms of the  
(anti) chiral superfield strengths ${\overline \cW}$ and $\cW$ which satisfy the Bianchi identities
\footnote{The derivatives $\cD^{ij}$ and ${\overline \cD}^{\, ij}$ are defined as 
$\cD^{ij}=\cD^{i\alpha}\cD^j_\alpha$ and
${\overline \cD}{}^{\,ij}={\overline \cD}{}^{\,i}_{\dot\alpha}{\overline
\cD}{}^{j{\dot\alpha}}$. See also appendix \ref{conventionsSUSY}.}
\be
\cD^{ij} \, \cW  \, = \,  {\overline \cD}{}^{\,ij} \, {\overline \cW} \ .
\label{N=2bi-i}
\ee
For an abelian gauge symmetry they can be solved by expressing the superfield strength in 
terms of an unconstrained prepotential $V_{ij}$:   
\be
\cW =  {\overline \cD}^{\, 4} \cD^{ij} \, V_{ij} \, , \qquad \quad 
{\overline \cW} = \cD^{\, 4} {\overline \cD}^{\,ij} \, V_{ij} \ .
\ee
The overall factors of ${\overline \cD}^{\, 4}$ and ${\cD}^{\, 4}$ guarantee that $\cW$ and ${\overline \cW}$ 
are chiral and anti-chiral, respectively, since ${\overline \cD}^{\, i}_\alpha {\overline \cD}^{\, 4} U=0$ for any
superfield $U$. 

Similarly to the ${\cal N}=0$ and ${\cal N}=1$ theories, duality transformations for ${\cal N}=2$ theories 
may be implemented \cite{Kuzenko:2000uh} in the path integral as a Legendre transform. One starts with 
the action 
\be
\label{Sinv}
\cS_{\text{inv}}=
\cS[\cW, {\overline \cW}]-\frac{\rm i}{8}\int d^8 \cZ\,\cW\cM
                                       +\frac{\rm i}{8}\int d^8\, {\overline \cZ}\, {\overline \cW}\, {\overline \cM}\,,
\ee
treating $\cW$ and ${\overline \cW}$ are  unconstrained superfields ({\it i.e.} not obeying the Bianchi identity). 
$\cM$ and its conjugate are determined by varying $S$ with respect 
to $\cW$ and ${\overline \cW}$
\be
{\rm i}\,{\cal M} \equiv 4\, \frac{\d }{\d \cW}\,
{\cal S}[\cW , {\overline \cW}]
 \, , \qquad \quad
- {\rm i}\,{\overline {\cal M}} \equiv 4\, 
\frac{\d }{\d {\overline \cW}}\, {\cal S}[\cW , {\overline \cW}]\,.
\label{N=2vd}
\ee
Their equations of motion 
\be
\cD^{ij} \, {\cal M}  \, = \,  {\overline \cD}^{\,ij} \, \overline {\cal M} 
\label{N=2em}
\ee
have the same functional form as the Bianchi identities (\ref{N=2bi-i}).

The infinitesimal duality transformations are therefore very similar to the 
${\cal N}=0$ and ${\cal N}=1$ ones:
\be
\d \cW  \, = \,  B \, {\cal M} \, , \qquad \quad
\d {\cal M}   \, = \,  -B \, \cW \, .
\label{N=2dt}
\ee
The requirement that $S_{\text{inv}}$ in eq.~(\ref{Sinv}) is invariant under this transformation
leads to the ${\cal N}=2$ analog of the duality conservation (NGZ) relation~(\ref{NGZbose}) 
\be
\int {\rm d}^8 \cZ\, 
\Big( \cW^2 + {\cal M}^2 \Big)  \, = \, 
\int {\rm d}^8 {\cZbar}\,
\Big( {\overline \cW}^2  +
{\overline {\cal M} }^2 \Big) \ ,
\label{N=2dualeq}
\ee
originally proven in \cite{Kuzenko:2000tg}, where it was called the ``$\NeqTwo$ self-duality equation''. 
This is the direct analog of the 
${\cal N}=1$ NGZ relation and reduces to it upon truncation of ${\cal N}=1$ chiral multiplet from 
the ${\cal N}=2$ vector multiplet and integration over two fermionic coordinates. One could in fact 
reconstruct the ${\cal N}=2$ constraint by starting from its ${\cal N}=1$ limit and requiring that it 
is manifestly supersymmetric and that it remains bilinear in superfields \cite{CKR}.

Solutions of this equation have proven fairly elusive.  The free ${\cal N}=2$ supersymmetric Maxwell action
 \be
{\cal S}_{\rm free} = 
\frac{1}{8}\int {\rm d}^8 \cZ \, \cW^2 +
\frac{1}{8}\int {\rm d}^8{\cZbar} \;{\cWbar}^2 
\label{N=2maxwell}
\ee
satisfies this constraint. An interacting action 
\be
{\cal S}  \, = \, 
\frac{1}{4}\int {\rm d}^8  \cZ \,  {\cal X} +
\frac{1}{4}\int {\rm d}^8{\overline \cZ} \,{\overline {\cal X}} \, ,
\label{N=2bi}
\ee
where the chiral superfield ${\cal X}$ is a functional 
of $\cW$ and $\overline \cW$ and is a solution of the constraint
\be
{\cal X}  \, = \,  {\cal X} \, {\overline \cD}^{\, 4} {\overline {\cal X}}  \, + \, 
\hf \, \cW^2 \, ,
\label{N=2con}
\ee
was proposed in~\cite{Ket1,Ket2}; this action was proven in \cite{Kuzenko:2000tg} to obey the 
$\NeqTwo$ self-duality constraint (\ref{N=2dualeq}).
This system may be solved perturbatively in the number of fields~\cite{Ket2,Kuzenko:2000uh, 
Ketov:2001dq,Bellucci:2001hd} and leads to an action of the form 
\be
\cS_{{\cal N}=2}= S_{\rm free} + \int {\rm d}^{12} {\cal Z} \; 
{\cal W}^2 \, {\overline{\cal W}}^2\,
{\cal Y}\; (\cD^4{\cal W}^2,\; {\overline \cD}^4{\overline {\cal W}}^2) + {\cal O}(\partial_\mu {\cal W})
\label{Ketov}
\ee
where ${\cal Y}$ is a Born-Infeld-type functional. The extra terms with space-time derivatives
$\partial_\mu {\cal W}$ are  required for ${\cal N}>1$ actions, see Appendix B.  
The system (\ref{N=2bi}), (\ref{N=2con}) was introduced in~\cite{Ket1,Ket2} as 
the ${\cal N}=2$ generalization of the Born-Infeld action.

An action exhibiting D3 brane type shift symmetry, exposing the spontaneous
breaking of translational invariance in the directions transverse to the brane,
was  proposed in \cite{Kuzenko:2000uh}. It simultaneously solves the ${\cal
N}=2$ NGZ constraint (\ref{N=2dualeq}).
\bea
\cS_{\rm BI} &=& \cS_{\rm free}  \, + \,  \cS_{\rm int}
\\
\cS_{\rm int} &=& 
{ 1 \over 8} \,  { \int {\rm d}^{12} {\cal Z}  }\, 
\Bigg\lbrace\cW^2\,{\overline \cW}^2\, \Bigg[ \lambda + 
\frac{\lambda^2}{2}\, \Big( \cD^4 \cW^2 + {\overline \cD}^{\, 4} {\overline \cW}^2 \Big)\label{5-int}  \non  \\
&~&~~~~+ \frac{\lambda^3}{4} \, 
\Big( (\cD^4 \cW^2)^2 + ({\overline \cD}^{\, 4} {\overline \cW}^2)^2 
+ 3\, (\cD^4 \cW^2)({\overline \cD}^{\, 4} {\overline \cW}^2)\Big)\Bigg]
 \non \\
&~&~~~~+{ 1 \over 3} \, 
\Bigg[ \frac{\lambda^{2}}{3} \cW^3 \Box {\overline \cW}^3 
+ \frac{\lambda^3}{2} \; \left(  (\cW^3 \Box {\overline \cW}^3) {\overline \cD}^{\, 4} {\overline \cW}^2 
+ ({\overline \cW}^3 \Box \cW^3) \cD^4 \cW^2 
+{ 1 \over 24} \cW^4 \Box^2 {\overline \cW}^4 \right) \Bigg] \non \\
&~&~~~~+ \, \cO(\cW^{10}) \Bigg\rbrace \, , 
\label{n2BIsoln}
\eea 
where we have introduced a dimensionful coupling $\lambda$, usually set to unity in the literature.
The unique term with no fermionic or space-time derivatives, $\cW^2\,{\overline \cW}^2$,
yields the known $F^4$ term of the Born-Infeld action, c.f. \app{appA}.  The sixth-order terms, apart from the 
$\cW^3 \Box {\overline \cW}^3$ terms with space-time derivatives, also correspond to a straightforward generalization of the bosonic BI model. 
This action was confirmed\footnote{Ref.~\cite{Bellucci:2001hd}  assigns $\Box$ a factor of $-\hf$ relative to the convention 
of \cite{Kuzenko:2000uh}.} in \cite{Bellucci:2001hd,Ketov:2001dq}; moreover, it belongs to the class of 
actions constructed in \cite{Bellucci:2001hd} which also exhibit another nonlinearly realized ${\cal N}=2$ 
supersymmetry algebra.

The construction we will detail in later sections will allow us to recover  
the action (\ref{n2BIsoln})  among infinitely many other actions.   While all
of these actions are expressible in the form of \eqn{N=2bi}, {\it i.e.}  given 
by the sum of a chiral and anti-chiral actions, they will not  generically 
satisfy  \eqn{N=2con}.  
Relaxing the requirement that the action has the form (\ref{N=2con}) leads, as we will see, to a variety of
actions with different properties.

\section{Construction of duality-satisfying actions}
\label{Sct:Action}

An important lesson from the construction of bosonic gauge theory duality-covariant actions 
was that the twisted self-duality constraint can be seen as more fundamental than the action.  Indeed, the twisted self-duality constraint determines the action
through the definition of the dual field $G$, see eq.~(\ref{Gdef}). The supersymmetric generalization 
of this feature is that an ${\cal N}=2$ twisted self-duality constraint should determine the action through 
the definition (\ref{N=2vd}) of the dual field $\cM$. 
We will discuss the ${\cal N}=2$ case in the same language as the generalized procedure~\cite{CKR} and so 
begin by describing the suitable twisted self-duality constraint and its deformation sources. We will then proceed, 
for a generic deformation source ${\cal I}$, to construct  an action with the desired properties. 

\subsection{Twisted self-duality.} 

As we saw in section~\ref{Sct:Review_Neq2}, $\cW$ and $\cM$ are interchanged by infinitesimal 
duality transformations. Similarly to the ${\cal N}=0$ case, they may be combined into $T$-variables 
which are simply rescaled by such transformations: there are two chiral  
\be
\label{def}
T^{+}= \cW- {\rm i} \; {\cal M}\, ,  \qquad  {\overline T}\,{}^{+} = \cW +  {\rm i} \; {\cal M}
\ee
and two anti-chiral fields 
\be
\label{def2}
T^{-}= \cWbar- {\rm i} \; \cMbar\, ,  \qquad {\overline T}\,{}^{-}= \cWbar  + {\rm i} \; \cMbar \ ,
\ee
each pair having one positive and one negative charge under the $U(1)$ duality transformation:
\be
\label{U1}
 \left(
                      \begin{array}{cc}
                 \delta T^+ \\
                 \delta  {\overline T}\,{}^{+} \\
                      \end{array}
                    \right)\ =\left(
                      \begin{array}{cc}
                      {\rm i}\; B&  0 \\
                        0 & -{\rm i}\;B \\
                      \end{array}
                    \right)  \left(
                      \begin{array}{cc}
                T^+ \\
                {\overline T}\,{}^{+} \\
                      \end{array}
                    \right) \, , \qquad \left(
                      \begin{array}{cc}
                \delta  T^- \\
                 \delta{\overline T}\,{}^{-} \\
                      \end{array}
                    \right)\ =\left(
                      \begin{array}{cc}
                      {\rm i}\;B&  0 \\
                        0 & -{\rm i}\;B \\
                      \end{array}
                    \right)  \left(
                      \begin{array}{cc}
                 T^- \\
              {\overline T}\,{}^{-} \\
                      \end{array}
                    \right) \, . 
\ee
\begin{table}[tp]%
\centering%
\begin{tabular*}{0.65\textwidth}{@{\extracolsep{\fill}} | c | c | c | }
 \hline
Combinations of superfields\ & Chirality & Charge  \\
  \hline
$T^{+}=\cW-{\rm i } {\cal M}$\ & $+$ & $+$  \\
  \hline
 ${\overline T}\,{}^{+}=\cW +{\rm i } {\cal M}$ & $+$ & $-$ \\
  \hline
$T^{-}=\cWbar-{\rm i } \cMbar $\ & $-$& $+$ \\
  \hline
${\overline T}\,{}^{-}= \cWbar+{\rm i } \cMbar $& $-$  &  $-$   \\
  \hline
\end{tabular*}
\caption{The four combinations of superfields have $\pm$ chirality and $\pm$ duality charge.}
\label{grav}
\end{table}
The behavior of these fields are summarized in \tab{grav}, which allows us to identify immediately 
the kind of superspace integral that is needed to turn some product of superfields into a supersymmetric 
action as well as identify its properties under duality transformations. The twisted self-duality constraint 
is the same as (\ref{linsd})
\be
 T^{+}= {\overline T}{}^{-}=0\, ,
\label{Neq2linsd}
\ee
 while the supersymmetric NGZ constraint (\ref{N=2dualeq}) becomes
\be
\int d^8 \cZbar\; {\overline T}\,{}^{+}\,T^+-  \int d^8   \cZ \; {\overline T}\,{}^{-}\,T^-=0 \ ,
\label{NGZ_Tform}
\ee
whose solutions we would like to construct.

As in the bosonic case, we begin with an ``initial source of deformation"  ${\cal I}(T^{-},{\overline T}\,{}^{+})$ 
which is a function of the superfields not set to zero by (\ref{Neq2linsd}).  
On dimensional grounds, any such ${\cal I}$ will depend 
on a dimensionful coupling $\lambda$. Moreover, since both chiral and anti-chiral 
superfields can appear as arguments, ${\cal I}$ is naturally a full superspace integral.
We will further assume that ${\cal I}$ is hermitian, which will lead to a simple 
characterization of the solutions to (\ref{NGZ_Tform}). 
The deformation of the linear 
twisted self-duality constraint (\ref{Neq2linsd}) will be given by
 \be
T_{}^+  = \frac{\delta {\cal I}(T^-,  {\overline T}\,{}^{+})}{\delta  {\overline T}\,{}^{+}_{}} \, , 
\qquad 
(T_{}^+)^*=  {\overline T}\,{}_{}^{-} = \frac{\delta {\cal I}(T^-,  {\overline T}\,{}^{+})}
{\delta  T^{-}_{}}\, ,
\label{BNBI}
\ee
where, in the second equality, we used the assumption that ${\overline{ {\cal I}(T^-, {\overline T}\,{}^{+})}} = 
{\cal I} (T^- , {\overline T}\,{}^{+})$. 
Any deformation source yields an action; the NGZ identity (\ref{NGZ_Tform}) identifies the deformation sources
leading to actions with duality-invariant equations of motion. Indeed, by replacing (\ref{BNBI}) into 
(\ref{NGZ_Tform}) we find a differential equation for ${\cal I}$:
\bea
0= \int d^8 \cZbar\; {\overline T}\,{}^{+} {\frac{\delta }{\delta  {\overline T}\,{}^{+}}} 
{\cal I}(T^-,  {\overline T}\,{}^{+}) -  \int d^8   \cZ \; 
T^- {\frac{\delta}{\delta  T^{-}}}{\cal I}(T^-,  {\overline T}\,{}^{+}) \, .
\label{NGZsdcpx1}
\eea
It is worth noting that, since ${\cal I}$ is a full superspace integral, each of the two superficially chiral integrals 
above is, in fact, also an integral over the full superspace. A notable difference from the  ${\cal N}=0$ case is that 
the supersymmetric NGZ constraint involves a space-time integral, which projects out possible total derivatives
in its integrand.
A solution to the equation~(\ref{NGZsdcpx1}) is that ${\cal I}$ is invariant under the $U(1)$ duality 
transformation~(\ref{U1}). Indeed, as in the bosonic case, the operator 
\be
\left({\overline T}^+ \frac{\delta }{ \delta {\overline T}{}^{+}}
-
{T}^- \frac{\delta }{ \delta {T}{}^{-}} \right)
\ee
``measures'' the charge under such transformations. A slightly more general
solution to \eqn{NGZsdcpx1} allows ${\cal I}$ to be invariant up to total derivatives. 

The twisted self-duality equations (\ref{BNBI}) can be solved recursively and yield $\cM$ as a series in 
$\lambda$ with coefficients which are functions of $\cW$ and ${\overline \cW}$ of appropriate degrees 
of homogeneity:
\be
\cM = \cM^{(0)}+\sum_{n\ge 1} \lambda^n\cM^{(n)}(\cW, {\overline\cW}) \ .
\label{Mseries}
\ee
Taking into account the fact that ${\cal I}$ is ${\cal O}(\lambda)$, the recursive solution for 
eqs.~(\ref{BNBI}) is
\be
\label{recurseM}
 \cM^{(n)} \equiv \lambda^{-n} \left( \frac{\delta}{\delta {\overline T}\,{}^{+}}  {\cal I} \left[T^-(\cW,\cM^{(n-1)}), {\overline T}\,{}^{+}(\cWbar,\cMbar^{(n-1)})\right] -\sum_{j=1}^{n-1} \lambda^j \cM^{(j)} \right) \, {\rm with}\, \, \lambda^{m > n}\rightarrow0\, ,
\ee
with $\cM^{(0)}$ being the solution to the linear twisted self-duality constraint, $\cM^{(0)}=-i\cW$.

With a solution in hand, the action may then be found by integrating the equations~(\ref{N=2vd}).  This 
can be done directly on a case by case basis.   We will show that it is in fact straightforward to carry out this integration for a general ${\cal I}$. The resulting action has a simple form as demonstrated by the examples detailed in the next section. 

As reviewed in section~\ref{Sct:Review_Neq2}, the NGZ consistency condition (\ref{N=2dualeq}), (\ref{NGZ_Tform})
is simply the requirement that the right-hand side of eq.~(\ref{Sinv}) -- with $\cW$, $\cM$ and their conjugates 
treated as independent fields -- remains invariant under duality transformations. While in general there exist many duality 
invariants, given an action $S[\cW, {\overline \cW}, \lambda]$, it is possible to construct a natural duality invariant 
expression\footnote{The overall sign is chosen for later convenience.}:
\be
\label{Sinv_ito_S}
\cS_{\text{inv}} = -{\lambda}\frac{d}{d\lambda} \cS[\cW, {\overline \cW}, \lambda] \ .
\ee
This construction is a particular example of the general fact (see e.g. \cite{Gaillard:1981rj,Kuzenko:2000uh}) that the derivative of a
duality-satisfying action with respect to a duality invariant parameter is duality-invariant.
To see that this is indeed the case let us carry out an infinitesimal duality transformation of this relation:
\be
\delta \cS_{\text{inv}}=-{\lambda}\frac{d}{d\lambda} \left(
\int d^8\cZ\; \delta \cW \frac{\delta \cS}{\delta \cW}
+
\int d^8{\overline\cZ}\;\delta {\overline \cW} \frac{\delta \cS}{\delta {\overline\cW}}\right) 
\ee
with $\delta \cW$ and $\delta {\overline \cW}$ given by (\ref{N=2dt}). The variation of the action with 
respect to $\cW$ and ${\overline \cW}$ can then be expressed,  using (\ref{N=2vd}), in terms of $\cM$ and 
its conjugate; thus, $\delta \cS_{\text{inv}}$ becomes
\be
\delta \cS_{\text{inv}}=-\frac{\rm i}{4}B\,{\lambda}\frac{d}{d\lambda} \left(
\int d^8\cZ\; \cM^2
-
\int d^8{\overline \cZ}\; {\overline \cM}^2\right) \ .
\ee
Since $\cW$ and ${\overline\cW}$ are independent of the coupling constant $\lambda$, we may freely add 
them to the parenthesis above:
\be
\delta \cS_{\text{inv}}=-\frac{\rm i}{4}B\,{\lambda}\frac{d}{d\lambda} \left(
\int d^8\cZ\; \left(\cW^2+\cM^2\right)
-
\int d^8{\overline \cZ}\; \left({\overline \cW}^2+{\overline \cM}^2\right)\right) \ .
\ee
This expression vanishes identically for $\cM$ and ${\overline \cM}$ satisfying the NGZ duality 
constraint~(\ref{N=2dualeq}), implying that (\ref{Sinv_ito_S}) indeed represents a valid choice of 
$\cS_{\text{inv}}$, though of course not the only possible one\footnote{For example, any function of $\lambda$ 
and $dS/d\lambda$ will lead to a possible candidate for the action. It is possible that all such choices are in 
fact equivalent through a change of initial deformation source, perhaps through a field redefinition.}. 

Clearly this choice of $S_{\text{inv}}$ allows us, through eq.~(\ref{Sinv}), to reconstruct the action 
in terms of $\cW, \,\cM$ and their conjugates. It is indeed easy to see that, upon use of  (\ref{Sinv_ito_S}), 
eq.~(\ref{Sinv}) becomes a first-order differential equation for $\cS[\cW, {\overline \cW}, \lambda]$ 
\be
-{\lambda}\frac{d}{d\lambda} \cS[\cW, {\overline \cW}, \lambda] =\cS[\cW, {\overline \cW}, \lambda]
-\frac{\rm i}{8}\int d^8 \cZ\;\cW\cM[\cW, {\overline \cW}, \lambda] 
+\frac{\rm i}{8}\int d^8 {\overline \cZ} \; {\overline \cW}{\overline \cM}[\cW, {\overline \cW}, \lambda] 
\ee
with the condition that at $\lambda=0$ its solution should be the free action. Its solution provides a form of the reconstructive identity for the action

\be
\cS= {{\rm i}\over 8 \lambda }  \int d \lambda   \Big [\int  {\rm d}^8 \cZ \, \cW {\cal M}[\cW, {\overline \cW}, \lambda]  
- \int {\rm d}^{8} \cZbar \,{\cWbar} \,\cMbar[\cW, {\overline \cW}, \lambda] \Big ]\,,
\label{niceaction}
\ee

\noindent where $\cM[\cW, {\overline \cW}, \lambda] $ and $\cMbar[\cW, {\overline \cW}, \lambda] $ are 
simultaneous solutions of the NGZ duality constraint (\ref{N=2dualeq})
and of the deformed twisted self-duality equation.  The latter introduces the dependence on the dimensionful 
coupling $\lambda$ through the initial deformation source. Given such a solution 
(\ref{Mseries}), (\ref{recurseM}), the action we are looking for is:
\be
\cS=   {\rm i} \int  {\rm d}^8 \cZ \,  \cW   \sum_{n=0}  \frac{ \lambda^n }{ 8(n+1)  }  
{\cal M}^{(n)}[\cW, {\overline \cW}]  + {\rm h.c.}\, .
\label{actionFromM}
\ee
One may then check, on a case by case basis, that both the dual field $\cM$ and its conjugate (also constructed from this action as in eq.~(\ref{N=2vd})) reproduce the dual field that was used to construct the action through eq.~(\ref{niceaction}).

This class of actions encompasses both the free action (obtained for ${\cal I}=0$) as well as the interacting 
actions reviewed in the previous section. The relevant $\cal X$ function is just
\be
{\cal X} \equiv  \frac{\rm i}{2 \lambda} \int {\rm d}\lambda\, \cW \, \cM \ .
\ee
This requires a very specific ${\cal I}_{BI}$ -- one that involves a superposition of an infinite number 
of initial source terms, much like the ${\cal N}=0$ Born-Infeld models discussed in~\cite{CKR,CKO}.   
We will present this aggregate deformation source through order $\lambda^3$ in \sect{Sct:Examples}.
It is not difficult to construct higher-order terms which reproduce the action (\ref{N=2bi})-(\ref{N=2con}). 

\subsection{Covariant construction of ${\cal N}=2$-SUSY duality-satisfying actions \label{summary_construction}}
Let us summarize here the ${\cal N}=2$ generalization of the bosonic covariant 
construction \cite{CKR,CKO} reviewed in section~\ref{Sct:Review_CKR} beginning with some 
initial deformation or counterterm $\delta \cS$. While some of the steps 
are very similar, others depart from the bosonic ones due mainly to the compact action 
introduced in eq.~(\ref{niceaction}).

\begin{enumerate}

\item Take the variation of the counterterm with respect to the field-strength superfield,  and 
express as a  function of $T^-$ and $\overline{T}{}^+$, 
\be 
\frac{\delta \Delta \cS}{\delta {\overline T}{}^{+}} \to 
\frac{\delta {\cal I}(T^-_{}, {\overline T}{}^{+ }, \lambda)}{ \delta {\overline T}{}^{+}}\,.
\ee

\item Introduce an ansatz for the deformation source ${\cal I}( T^{-}_{}, {\overline T}{}^{+}, \lambda)$. 
In general, this may be taken to depend on all possible duality invariants.

\item Constrain the dual field strength to this variation: 
\be
\label{defGen}
T^+_{}=\frac{\delta {\cal I}( T^{-}_{}, {\overline T}{}^{+}, \lambda)}{ \delta {\overline T}{}^{+}}
\ee

\item \label{step:solveM} 
Solve \eqn{defGen} iteratively for the dual field $\cM=\cM[\cW, {\overline \cW}, \lambda]$ 
and its conjugate while checking that 
the NGZ duality constraint is satisfied. Any $U(1)$-invariant hermitian deformation source will 
automatically lead to solutions which pass this test.

\item \label{step:action} Use $\cM$ and its conjugate found at step~\ref{step:solveM} to construct the action 
\be
\cS= {{\rm i}\over 8 \lambda }  \int d \lambda   \Big [\int  {\rm d}^8 \cZ \, \cW \, {\cal M}[\cW, {\overline\cW},\lambda] 
- \int {\rm d}^{8} \cZbar \,{\cWbar} \cMbar[\cW, {\overline\cW},\lambda]\Big ]
\ee
while checking for additional desired properties and enlarging the ansatz for ${\cal I}$ if necessary.

\item \label{verify} Verify that $\cM$ and its conjugate used at step~\ref{step:action} are reproduced as
\be
{\rm i}\,{\cal M}[\cW, {\overline \cW}, \lambda]  \equiv 4\, \frac{\d }{\d \cW}\,
{\cal S}[\cW , {\overline \cW}]
 \, , \qquad \quad
- {\rm i}\,{\overline {\cal M}}[\cW, {\overline \cW}, \lambda]  \equiv 4\, 
\frac{\d }{\d {\overline \cW}}\, {\cal S}[\cW , {\overline \cW}]
\ee

\end{enumerate}

\noindent
It is important to point out that the last step above is not a substitute for any of the earlier steps.
For example, it is possible to construct deformation sources ${\cal I}$ which, while not solving the NGZ 
constraint  lead nevertheless through \eqn{defGen} to dual fields $\cM$ and an action which reproduces 
them. 

We will proceed in the next section to apply this construction to recover the actions
reviewed in section~\ref{Sct:Review_Neq2} as well as  duality covariant actions with manifest ${\cal N}=2$ supersymmetry and quite novel structure.


\section{Duality examples }
\label{Sct:Examples}

In this section we will follow the steps outlined above and discuss four
examples of deformation sources and their corresponding duality-covariant
actions. We present each source with an arbitrary scalar pre-factor  $(a,b,c,d)$
.  In each subsection we will mention the relevant value necessary to match
terms present in the BI action given in \eqn{n2BIsoln}.
At higher orders in the number of fields, the nested super derivatives can
become quite lengthy.  To shorten the expressions we introduce the notation
\begin{eqnarray}
{\cH{n}}&=&\SD( \cW^2 \cHb{n-1})\,\\
{\cHb{n}}&=&\SDbar(  \cWbar^2 \cH{n-1})\, ,
\label{Hnotation}
\end{eqnarray}
with $\cH{0} \equiv \SD( \cW^2)$ and $\cHb{0} \equiv \SDbar({\cWbar^2})$. This
notation will also eliminate the explicit space-time derivatives, unless they
appear already in the deformation source.

\subsection{$ (T^-)^2 ({\overline T}\,{}^{+})^2$}
\label{sec:I2}
The lowest dimension ``initial source of deformation'' which is manifestly invariant under duality 
transformations is
\be \label{I2}
{\cal I}_1= a\,{\lambda} \int {\rm d}^{12} \cZ \; (T^-)^2 ({\overline T}\,{}^{+})^2 \ ;
\ee
it also has a direct counterpart in the bosonic theory. Upon use of the definition of $T^\pm$ and its 
conjugate in terms of $\cW$ and $\cM$,  \eqn{BNBI} can be written as:
\be
{\cal M}= -{\rm i} \,  \cW  +  {2 a \lambda\,{\rm i}}  \, \Big (\SDbar (\cWbar- {\rm i}  \cMbar)^2\Big ) 
(\cW +  {\rm i} \, {\cal M})\,.
\label{sol}
\ee
This equation, and its conjugate involving $\cMbar$, can be solved recursively order by order in
$ \lambda$. The solution is relatively compact\footnote{It reflects the fact that when the source of deformation 
is a single quartic term, we have a cubic deformation of the linear constraint, as was also noticed in the 
``Model A'' explored in \cite{CKO}.} with the coefficients of $\cM = \sum_{n} \lambda^{n} \cM^{(n)}$ being 
given by
\bea
  \cM^{(0)}&=& - {\rm i}\; \cW \,,\\
\cM^{(n)}|_{n>0} &=& (- 2)^{5-n} a \sum_{l=0}^{n-1} \sum_{q=0}^{n-(1+l)} \alpha(l,q;n)
  \SDbar  [ \, \cMbar{}^{(n-(1+q+l))} \cMbar{}^{(q)} \cM^{(l)} \,  ]\, 
\label{recursiverelation}
\eea
with 
\bea
\alpha(q,l;n) &\equiv& \xi_2(q)  \xi_2(l)  \xi_2 (n-l-q-1)\, .\\
\xi_2(x)|_{x>0}&\equiv&  (-2)^{x}/2\ ,\\
\xi_2(x)|_{x=0}&\equiv& 1\ .
\eea

With the notation introduced in \eqn{Hnotation}, the first few terms in the expansion of $\cM$ 
are
\be
{\cal M}= -{\rm i} \,  \cW
+ 16 a\,{\rm i}  \, \lambda \cW \cHb{0}   -  \frac{{\rm i}}{2} (16 a)^2 \lambda^2 \; \cW\;   \left(
   (  \cHb{0}  )^2 + 2 \cHb{1}  \right)+ \cdots \ .
\ee
The action is then given directly by \eqn{actionFromM}; since it is linear in $\cM$,  that a recursive 
solution for $\cM$  automatically translates into a recursive expression for the action. 
Here we choose to solve the recursion and express it through $\lambda^4$ in the form of a Hermitian action:
\begin{multline}
\cS_1^{\rm int}= \int {\rm d}^{12} \cZ \cW^2 \cWbar^2 \Bigg\{-2 a \lambda 
+16 a^2 \lambda ^2  (\cH{0}+\cHb{0})\\
-128 a^3 \lambda ^3 \left(\cH{0}^2+2 \cH{1}+\cHb{0}^2+2 \cHb{1}\right)\\
+1024 a^4 \lambda ^4  \left(6 (\cH{0} \cH{1}+\cHb{0} \cHb{1})+4 (\cH{2}+\cHb{2})+\cH{0}^3+\cHb{0}^3\right)\\
-8192 a^5 \lambda ^5 \Bigg(6 \Big(\cH{0}^2 \cHb{0}^2+2 (\cH{0} \cH{2}+\cHb{0} \cHb{2})+2 \cH{1}^2+2 \cHb{1}^2\Big)\\
+8 \left(\cH{1} \cH{0}^2+\cH{3}+\cHb{0}^2 \cHb{1}+\cHb{3}\right)+\cH{0}^4+\cHb{0}^4\Bigg)
+ {\cal O}(\lambda^6) \Bigg\} \ .
\label{ExampleAAction}
\end{multline}

Note that setting $a=-2^{-4}$ recovers the terms in the BI action, \eqn{n2BIsoln} through $\lambda^2$ which 
do not contain the space-time Laplacian, as well as relevant contributions at higher 
 orders. While some sequences of terms -- such as those depending only on 
powers of $\cH{0}$ -- can be resummed, it does not appear that this action has a closed-form expression.

\subsection{$(T^-)^3 \Box ({\overline T}\,{}^{+})^3$}

As reviewed in section~\ref{Sct:Review_Neq2}, the term $(\cW^3 \Box\, {\overline
\cW}^3)$ in the  BI action (\ref{n2BIsoln}) is required if that action is to be
interpreted as the supersymmetric D3-brane action.  Such a term will not appear
in an action of the type (\ref{actionFromM}) unless we add a term like $(T^-)^3
\Box ({\overline T}\,{}^{+})^3$ to the initial deformation source.  Such terms
may be obtained from those discussed in \cite{CKR} by dressing them with
space-time derivatives.
Let us therefore consider the duality-invariant 
\be \label{I3}
{\cal I}_{2}= b \lambda^2  \int {\rm d}^{12} z \; (T^-)^3 \Box ({\overline T}\,{}^{+})^3 \ .
\ee
The twisted self-duality equation \eqn{BNBI} can be written as
\be
{\cal M}= -{\rm i}\,  \cW  +3\,{\rm i}\, b\,  \lambda^2  \, (\cW +  {\rm i}\, {\cal M})^2  \; \Big (\SDbar \, \Box  (\cWbar-{\rm i } \cMbar)^3\Big ) 
\label{sol2}
\ee
and, together with its conjugate, can be solved for $\cM$ and ${\overline \cM}$. The first few terms in 
their solution are
\begin{multline}
{\cal M}= -{\rm i}\,  \cW
+ 2^5 \times 3 \, {\rm i}\,  b \; \lambda^2  \; \cW^2 \; (\SDbar \, \Box \,  \cWbar^3\Big ) \\
-2^{9} \times 3^2 \, {\rm i}\,  b^2 \; \lambda^4 \;   \cW^2 \Big[
2 \cW  \;  \Big(\SDbar \, \Box \,  \cWbar^3\Big )^2 
+ 3 \; \SDbar  ( \Box \,  (  \cWbar^4  \SD \bar \cW^3   )   \Big] \\
+2^{13} \times3^3 \, {\rm i}\, b^3 \;\lambda ^6 \cW^2  \Big[
12 \cW \SDbar\left(\square \left(\cWbar^4 \mathcal{D}^4\left(\square \left(\cW^3\right)\right)\right)\right) \SDbar\left(\square \left(\cWbar^3\right)\right)\\
+9 \left(\SDbar\left(\square \left(\cWbar^5 \mathcal{D}^4\left(\square \left(\cW^3\right)\right)^2\right)\right)
+\SDbar\left(\square \left(\cWbar^4 \mathcal{D}^4\left(\square \left(\cW^4 \SDbar\left(\square \left(\cWbar^3\right)\right)\right)\right)\right)\right)\right)\\
+5 \cW^2 \SDbar\left(\square \left(\cWbar^3\right)\right)^3\Big]
+\cdots \ .
\end{multline}
Due to the presence of the Laplacian, it is inconvenient to use the notation (\ref{Hnotation}) 
for this deformation; we will instead express the action in terms of $\cW$ and its conjugate. 
Through order $\lambda^4$ it is
\begin{multline}
\cS_{2}^{\rm int}= \int {\rm d}^{12} \cZ \Bigg\{
-4 b \lambda ^2 \left(\cW^3 \square \left(\cWbar^3\right)+\cWbar^3 \square \left(\cW^3\right)\right)\\
+2^6 {\frac{9}{5} }b^2 \lambda ^4 \Big(3 \cW^3 \square \left(\cWbar^4 \SD\left(\square \left(\cW^3\right)\right)\right)+2 \cWbar^4 \square \left(\cW^3\right) \SD\left(\square \left(\cW^3\right)\right)\\
+2 \cW^4 \square \left(\cWbar^3\right) \SDbar\left(\square \left(\cWbar^3\right)\right)+3 \cWbar^3 \square \left(\cW^4 \SDbar\left(\square \left(\cWbar^3\right)\right)\right)\Big) +\cdots \Bigg\} \ .
\end{multline}  
Here we chose the interaction terms in the action as a full superspace integral, making use of the fact that the 
nonlinear terms in $\cM$ contain the appropriate chiral projector.
It is not difficult to see that, by choosing $b = - \frac{1}{576} = - 2^{-6}\; 3^{-2}$ we recover the  $\Box$-dependent
 term in the $\lambda^2$ contribution to the BI action  \eqn{n2BIsoln} as well as relevant contributions at higher 
 orders.

\subsection{$(T^-)^4 \Box^2 ({\overline T}\,{}^{+})^4$}

Invariants of type \eqn{I2} naturally generalize to higher orders -- for example
\be \label{I4}
{\cal I}_{3}= c {\lambda^3 } \int {\rm d}^{12} z \; (T^-)^4 \Box^2 ({\overline T}\,{}^{+})^4 \ .
\ee
With this deformation source, the twisted self-duality equation \eqn{BNBI} is
\be
{\cal M}= -{\rm i}\,  \cW  +  4  \,{\rm i}\, c\, {\lambda^3 } \;   \; (\cW +  {\rm i}\, {\cal M})^3  \; \Big (\SDbar \, \Box^2  (\cWbar-{\rm i } \cMbar)^4\Big ) \ .
\label{sol4}
\ee
As in the previous two cases, this equation and its conjugate can be solved recursively for $\cM$ 
and ${\overline\cM}$, leading to
\begin{multline}
{\cal M}= -{\rm i}\,  \cW + 2^{9} \,{\rm i}\, c 
   \lambda ^3 \cW^3 \SDbar (\square^2
    (\cWbar^4 ) )\\
   -2^{17} \,{\rm i}\, c ^2
   \lambda ^6 \cW^3  (4 \SDbar   (\Box^2  (\cWbar^6 \SD (\square^2
    (\cW^4 ) ) ) )
      +3
   \cW^2 \SDbar (\square^2     (\cWbar^4 ) )^2 )\\
   +2^{26} \,{\rm i}\, c ^3 \lambda ^9 \cW^3 \Big(12 \cW^2
   \SDbar (\square^2    (\cWbar^4 ) ) )
   \SDbar (\square^2  (\cWbar^6
   \SD (\square^2
    (\cW^4 ) ) ) )\\
   +9   \SDbar (\square^2  (\cWbar^8
   \SD (\square^2
    (\cW^4 ) )^2 ) )\\
   +8
   \SDbar (\square^2  (\cWbar^6
   \SD (\square^2  (\cW^6
   \SDbar (\square^2
    (\cWbar^4 ) ) )
   + 6 \cW^4 \SDbar (\square^2
  (\cWbar^4 )  )^3\Big)\cdots \ .
 \end{multline}
Absorbing  the overall chiral projector in the nonlinear terms into the integration measure and thus expressing 
the action as a hermitian full superspace integral we find, through ${\cal O}(\lambda^6)$ that
\begin{multline}
\cS_{3}^{\rm int}= \int {\rm d}^{12} \cZ \Bigg\{-16 c \lambda ^3 \left(\cW^4 \square \left(\square \left(\cWbar^4\right)\right)+\cWbar^4 \square \left(\square \left(\cW^4\right)\right)\right)\\
+
2^{14}{\textstyle \frac{1}{7}} c^2 \lambda ^6 \Big(4 \cW^4 \square \left(\square \left(\cWbar^6 \SD\left(\square \left(\square \left(\cW^4\right)\right)\right)\right)\right)+3 \cWbar^6 \square \left(\square \left(\cW^4\right)\right) \SD\left(\square \left(\square \left(\cW^4\right)\right)\right)\\
+3 \cW^6 \square \left(\square \left(\cWbar^4\right)\right) \SDbar\left(\square \left(\square \left(\cWbar^4\right)\right)\right)\\
+4 \cWbar^4 \square \left(\square \left(\cW^6 \SDbar\left(\square \left(\square \left(\cWbar^4\right)\right)\right)\right)\right)\Big)+\cdots \Bigg\} \ .
\end{multline}
To recover the  $\Box^2$  term in the $\lambda^3$ contribution to the BI action  \eqn{n2BIsoln}, as well as 
relevant terms at higher orders, we can set $c= -\;  2^{-12}\; 3^{-2}$.

\subsection{$(T^-)^2 ({\overline T}\,{}^{+})^2 \SDbar((T^-)^2) \SD(  ({\overline T}\,{}^{+})^2) $}

Another interesting invariant constructed out of four $T^-$ and four ${\overline T}\,{}^{+}$ factors and a number 
of super-derivatives is:
\be
\label{I4b}
{\cal I}_{4}= {\lambda^3 d } \int {\rm d}^{12} \cZ (T^-)^2 ({\overline T}\,{}^{+})^2 
               \SDbar((T^-)^2) \SD(  ({\overline T}\,{}^{+})^2) \ .
\ee
The resulting constraint equation is:
\begin{multline}
{\cal M}= -{\rm i}\,  \cW  + 2\,{\rm i} \, d \; \lambda ^3  (\cW+{\rm i}\, \cM) 
\Big\{ \SDbar\left(\left(\cWbar-{\rm i}\, \cMbar\right)^2\right) 
\SDbar\left(\left(\cWbar-{\rm i}\,  \cMbar\right)^2 \SD\left((\cW+{\rm i}\,  \cM)^2\right)\right)\\
+ \SDbar\left(\left(\cWbar-{\rm i}\,  \cMbar\right)^2 \SD\left((\cW+{\rm i}\,  \cM)^2 
\SDbar\left(\left(\cWbar-{\rm i}\,  \cMbar\right)^2\right)\right)\right)\Big\} \ .
\end{multline}
Solving it recursively leads, order by order in $\lambda$, to the following interaction terms in the action:
 \begin{multline}
\cS_{4}^{\rm int}= \int {\rm d}^{12} \cZ  \, \cW^2 \cWbar^2 \Bigg\{
-16 d \lambda ^3 (\cH{1}+\cHb{1})+
4096 \; d^2\; \lambda ^6  \Big(2 (\cH{1} \cH{2}+\cHb{1} \cHb{2})+\\
\cH{0} \cH{1}^2+\cH{4}+\cHb{0} \cHb{1}^2+\cHb{4} \Bigg)
+{\cal O}(\lambda^9) \Bigg\}\ .
\end{multline}

The leading $\lambda^3$ term recovers the final $\lambda^3$ terms in the BI action (\ref{n2BIsoln}) with $d$ 
 chosen to be $d=2^{-10}$. 

\subsection{BI action through $\lambda^3$} 
 
By suitably combining the deformation sources discussed above with the mentioned relative coefficients 
\begin{multline}
{\cal I}_{BI}=-\int {\rm d}^{12} \cZ \Bigg(
\lambda\; 2^{-4} \; (T^-)^2 ({\overline T}\,{}^{+})^2
+  \lambda^2\; 2^{-6}\;3^{-2}\; (T^-)^3 \Box ({\overline T}\,{}^{+})^3\\
 + \lambda^3\;  2^{-12} 3^{-2} (T^-)^4 \Box^2 ({\overline T}\,{}^{+})^4
- \lambda^3 \; 2^{-10} \; (T^-)^2 ({\overline T}\,{}^{+})^2 \SDbar((T^-)^2) \SD(  ({\overline T}\,{}^{+})^2) \\
+{\cal O}(\lambda^4)\Bigg)\,
\end{multline}
and carrying out the procedure of section~\ref{summary_construction} 
we recover the Born-Infeld action from the literature given above in \eqn{n2BIsoln}.  While the solution to 
the twisted self-duality constraint is inherently nonlinear in the initial deformation source, to this order 
only a small number of cross terms are actually relevant. 
At higher orders in $\lambda$ more such terms will become important together with the appearance of new 
invariants that may be added to the initial deformation source. However, not all cross terms can be modified 
by adding higher order invariants. For example, at $\lambda^4$ the only new source should likely be 
$ (T^-)^5 \Box^3 ({\overline T}\,{}^{+})^5$ which clearly cannot modify the majority of ${\cal O}(\lambda^4)$ terms in the action.

\subsection{More general models} 

Any $U(1)$ duality invariant source can be used in this procedure, and the new
form of the action makes it rather trivial to combine various sources to tailor
craft  duality-consistent ${\cal N}=2$ actions incorporating such sources --
such as recovering \eqn{n2BIsoln}.  Consider for example:
\bea
 \label{new1S}
 {\cal I}_{\rm gen~A}^{(n,m)}&=&\lambda_{(n,m)}  (\Box^n  {\overline T}{}^{(+)}{}^2) \Box^m (\Box^n {T}{}^{(-)}{}^2)\,,\\
 {\cal I}_{\rm gen~A'}^{(n,m)}&=&\lambda_{(n,m)}  (\Box^n \SD {\overline T}{}^{(+)}{}^2)  \Box^m (\Box^n \SDbar {T}{}^{(-)}{}^2)\,,\\
 {\cal I}_{\rm gen~A''}^{(n,m)}&=&\lambda_{(n,m)}  (\Box^n \partial^\mu{\overline T}{}^{(+)}\partial^\nu{\overline T}{}^{(+)})  \Box^m (\Box^n \partial_\mu{T}{}^{(-)}\partial_\nu{T}{}^{(-)})\,,\\
 {\cal I}_{\rm gen~B}^{(n,m)}&=&\lambda_{(n,m)}^2  (\Box^n  {\overline T}{}^{(+)}{}^3) \Box^m (\Box^n {T}{}^{(-)}{}^3)\,,\\
 {\cal I}_{\rm gen~B'}^{(n,m)}&=&\lambda_{(n,m)}^2  (\Box^n \SD {\overline T}{}^{(+)}{}^3) \Box^m (\Box^n \SDbar {T}{}^{(-)}{}^3)\,,\\
 \label{new6S}
  {\cal I}_{\rm gen~C}^{(n_1,n_2,m)}&=&\lambda_{(n_1,n_2,m)}^3  (\Box^{n_1}  {\overline T}{}^{(+)}{}^2) (\Box^{n_2}  {\overline T}{}^{(+)}{}^2  ) \Box^m (\Box^{n_1} {T}{}^{(-)}{}^2)(\Box^{n_2} {T}{}^{(-)}{}^2)\,,
\eea
for $n_i,m \ge 0$.   It is important to note that the $\lambda_{n_i,m}$ above
will have different dimensions based upon the value of $n_i,m$.   Each of these
is a duality invariant initial source, and generates novel duality-covariant
actions through the procedure specified above.  It is not difficult to see how
these patterns generalize to an infinite set of other initial sources.

The BI action is non-renormalizable by power counting. As a test of the
preservation of duality symmetries at the quantum level one may  construct
counterterms in the BI model and check whether they preserve the classical
$U(1)$ duality symmetry as well as whether the higher-order terms generated by
our procedure reproduce those obtained by direct calculation. Compared to
supergravity theories, the simplicity of the BI model provides a clear advantage
as a testing ground for such questions.  One-loop calculations in the $\NeqTwo$
BI theory have been already carried out in \cite{Shmakova:1999ai,
DeGiovanni:1999hr} where it was found that the relevant momentum space on-shell
counterterm is
\bea
\label{Gamma1}
\Gamma_1^{\text{div}} = c(\epsilon)\int dp_1dp_2dp_3dp_4\delta^4\left(\sum
p_i\right)\int d^8\theta
\left(s^2+\frac{4}{3}t^2\right) \cW(p_1)\cW(p_2){\overline\cW}(p_3){\overline\cW}(p_4) \ .
\eea
Here $c(\epsilon)$ is a divergent coefficient. This counterterm may be written
in position space in several ways, related by use of on-shell conditions
$p_i^2=0$. One way, chosen in \cite{DeGiovanni:1999hr} places one derivative on
each superfield factor: 
\bea
\Gamma_1^{\text{div}} = c(\epsilon)\int d^4x d^8\theta
\left( \partial^\mu\cW\partial_\mu \cW \partial^\nu{\overline\cW}\partial_\nu{\overline\cW}
+\frac{4}{3} \partial^\mu\cW \partial^\nu\cW \partial_\mu{\overline\cW} \partial_\nu{\overline\cW} \right)\ .
\eea
Using the on shell conditions, $p_i^2=0$, the first term may also be written as 
\be
\int d^4x d^8\theta \;\Box(\cW^2)\Box({\overline\cW}^2) \ ,
\ee
which identifies it as as arising from ${\cal I}_{\text{gen}~A}^{(1,0)}$. The
second term, which is similar to  bosonic terms considered in \cite{CKO}, arises
from ${\cal I}_{\text{gen}~A''}^{(0,0)}$.
Through our procedure, these deformation sources make definite predictions about
some of the higher order terms that should appear in perturbative higher loop
calculations. In particular, these terms will necessarily be accompanied by
higher powers of the coefficient $c(\epsilon)$ in eq.~(\ref{Gamma1}) thus
implying, apart from the field dependence, also a definite strength of the
corresponding divergence.
It should be interesting to check explicitly whether these predicted higher
order terms correspond to  the results of multi-loop and multi-leg perturbative
calculations and, if they do not, whether the deformation source may be suitably
modified to accommodate the difference. 
New deformation sources will, however, always be necessary at each loop order.
Indeed, the nonlinearity of eq.~(\ref{defGen}) implies that all higher-order
terms which are generated will contain more fields than the deformation source.
Consequently, new deformation sources will be required to all orders in
perturbation theory at least for the four-superfield counterterm.

As we explained above, as long as the action depends only on a chiral and
anti-chiral superfields $\cW$ , $\cWbar$ and their spinorial and space-time
derivatives, one is free to make any choice of  ${\cal I}$. When one such choice
is made \eqn{BNBI} supplies a recursive procedure for obtaining ${\cal M}$.
While it may not always be as simple as the quartic deformation given in
\eqn{I2} the method described above, generalized from \cite{CKR}, and as
demonstrated also in \cite{CKO}, allows one to recursively produce the action to
any desired level. At each order the number of invariants that can be used is
limited, given a fixed engineering dimension of the coupling constant.  We see,
therefore, choices made at lower orders typically have definite consequences at
higher orders in the dimensional coupling.

It is in principle possible to impose additional requirements of our
construction, in particular the existence of additional symmetries beyond
duality. In general however, it is not immediately clear how to encode such
requirements in the choice of initial deformation source. It may be possible to
first identify the properties of $\cM[\cW, {\overline\cW},\lambda]$ from the
action (\ref{niceaction}) and then require that the twisted self-duality
equation is also invariant under the same transformations.  This could prove too
strong a requirement, as many algebraic equations can have the same solution.
Alternatively, one may start with the most general deformation source with
arbitrary coefficients and determine them by requiring that the resulting action
is invariant under the desired symmetries.
An important example in this direction is the construction of the action in
\cite{Kuzenko:2000uh} -- which we have reproduced with our method -- where in
addition to duality symmetry the action has to satisfy a certain constant shift
symmetry\footnote{The shift symmetry of this type is reminiscent of the shift
symmetry part of the $E_{7(7)}$ in ${\cal N}=8$ supergravity, acting on scalar
fields.}, associated with the D3 brane action: 
\be
\delta \cW= \sigma + {\cal O} (\cW,  \cWbar) \ .
\ee
Similarly, in \cite{Bellucci:2001hd} the $\NeqTwo$ supersymmetric Born-Infeld  action was 
required, apart from duality invariance, to also have a partially broken ${\cal N}=4$ supersymmetry.
It is, however, not obvious that the construction of the action is algorithmic and whether actions of a 
different structure may be obtained by relaxing any one of these properties. 
Given any self-dual hermitian function ${\cal I}$ of $\cW$, ${\overline\cW}$ and their derivatives, our construction
directly constructs, order by order, an action which is self-dual and covariant, thus showing that there exist infinitely 
many solutions to the self-duality constraints. Further symmetry requirements may also be imposed by a 
suitable choice of deformation source ${\cal I}$.

\section{Discussion}
\label{Sct:Discussion}

In this article we explore the space of $U(1)$-duality invariant actions with
rigid $\cN=2$ supersymmetry employing and extending the methods developed and
explored by three of the current authors in the previous publication \cite{CKR}.
For these models we identify a useful presentation of the action and the $U(1)$
duality constraint such that---when the latter is solved perturbatively---the
construction of the former follows immediately. Namely, when a certain choice of
the manifestly duality invariant source of the deformation of the linear twisted
self-duality constraint is made, ${\cal I} (T, {\overline T}, \lambda)$,  its
derivative provides a dual superfield $\cM=\cM(\cW, {\overline \cW}, \lambda)$
as  the function of the original superfields $\cW, {\overline \cW}$ and as a
power series in the  coupling $\lambda$, as one can see from eqs. (\ref{BNBI})
and (\ref{Mseries}). The chiral part of the action  is reconstructed by the
integration over $\lambda$ of the product $ \cW  {\cal M}(\cW, {\overline \cW},
\lambda)$,  and the conjugate to it provides the anti-chiral part of the action,
as shown in eq. (\ref{niceaction}). 

Employing this approach, we have identified several initial deformations, which,
after applying the recursive method, collectively reproduce the 
actions found in \cite{Ket1,Ket2,Kuzenko:2000tg,Kuzenko:2000uh}.  Beyond those, there
are further classes of deformations, which lead to a rich variety of
duality-invariant models with $\cN=2$ supersymmetry. 
One important observation has been described in \cite{CKR} for the
non-supersymmetric and ${\cal N}=1$ supersymmetric theories: when the initial
source of deformation is quartic in $F$, the deformation of the linear twisted
self-duality condition leads to a {\it Born-Infeld type} action, that is an action
containing all powers of $F$ up to infinity. The higher order terms are
necessary to maintain the duality invariance of the equations of motion order by
order. Using the corresponding construction for $\cN=2$ supersymmetric models
detailed in this paper leads to the same conclusion. 

The next natural step towards understanding the implications of $E_{7(7)}$ duality 
for the UV behavior of $\NeqEight$ supergravity is the construction of simpler examples, 
such as ${\rm U}(1)$ duality-consistent deformed ${\cal N}=2$ supergravity theories.
While the jump from rigid-supersymmetry to supergravity is non-trivial, we expect 
${\cal N}=2$ supergravity to be an excellent proving ground.
It is possible that a suitable covariantization  of the twisted self-duality constraint 
(\ref{defGen}) coupled with an appropriate choice of deformation source will allow us to 
construct the analog of the actions discussed in this paper in the presence of local 
$\NeqTwo$ supersymmetry.

We expect \cite{us} that a suitable covariantization  of the twisted self-duality constraint 
(\ref{defGen}) coupled with an appropriate choice of deformation source will allow us to 
construct the analog of the actions discussed in this paper in the presence of local 
$\NeqTwo$ supersymmetry.

Let us briefly comment on the possible consequences of such covariant
constructions for UV divergences in $\NeqEight$ supergravity.  If existent, the
first UV divergence for a four-point amplitude (which is sufficient to consider)
will be the $\cN=8$ supersymmetric completion of a term of the form
$f(s,t,u)R^4$, which necessarily contains a local quartic term of the form
$f(s,t,u)(dF)^4$ in momentum space. Here $f$ is a polynomial of the usual
Mandelstam variables whose degree depends on the loop order at which the
divergence arises.

Assuming $E_{7(7)}$-symmetry to persist unmodified at the quantum level, there
exist $E_{7(7)}$-invariant counterterms at sufficiently high loop order, as
originally shown in \cite{Kallosh:1980fi} and more recently reinforced in
\cite{Beisert:2010jx}.
The sufficiency of this reasoning was questioned in \cite{Kallosh:2011dp}: as
$E_{7(7)}$ is a continuous global symmetry, it requires the conservation of the
corresponding NGZ current\cite{Gaillard:1981rj} in addition to the
$E_{7(7)}$-invariance of the counterterm candidates. 
It was, however, subsequently suggested in \cite{BN} that there exists a procedure 
of perturbative deformation of the linear self-duality constraint which always allows the addition to
the action of the candidate counterterm along with all other higher
terms required for the conservation of the NGZ current. 

The covariant procedure introduced in \cite{BN} required generalization to recover the venerable 
bosonic Born-Infeld action~\cite{CKR}.  This class of generalizations has been applied to consider higher
derivative terms~\cite{CKO} and in this paper we have presented an application to ${\cal N}=2$ global supersymmetry. 
Discovering the way to meaningfully apply such generalizations to $\cN=8$ supergravity could shed light on the 
UV-finiteness question.  In principle, one would start from the
classical $\NeqEight$ theory, in which the supersymmetrization of the Ricci
scalar contains terms quadratic in vector fields: 
\be
 \cS_{\cN=8}(g_{\mu\nu}, F_{\mu\nu}, ...; \kappa^2)=\int{1\over 2\kappa^2} (R- F{\cal N(\phi)} F+...) \ .
\label{cl} 
\ee
One could imagine that attempting to construct a Born-Infeld type $\cN=8$ supergravity, by adding suitable
deformation sources and applying the covariant duality construction, results in two possible scenarios:

\begin{itemize}

\item construction of an $\NeqEight$ Born-Infeld type supergravity is possible, either for a general value 
$g$ of the four-vector terms 
     \be
      \cS_{\NeqEight}^{BI}(g_{\mu\nu}, F_{\mu\nu}, ...; \kappa^2, g^2)
      =\int{1\over 2\kappa^2} (R- F{\cal N(\phi)} F+\cdots) + g^{2} F^4 f_4(s,t,u) +
       \cdots+g^{2m}F^n f_n(s, t, u,...)+\cdots 
     \label{BIs} 
     \ee
or only for specific coefficients $g_n(\kappa)$ of the $n$--vector terms
      \begin{multline}
      \cS_{\NeqEight}^{BI'}(g_{\mu\nu}, F_{\mu\nu}, ...; \kappa^2, g^2)
      =\int{1\over 2\kappa^2} (R- F{\cal N(\phi)} F+\cdots) 
      + g_4(\k^2) F^4
      f_4(s,t,u) + 
      \cdots\\
      +g_{m}(\k^2)F^n f_n(s, t, u,...)+\cdots
     \label{BI2s} 
    \end{multline}
It is however only this second option which has the possibility of being consistent with perturbation theory of the 
undeformed $\NeqEight$ supergravity\footnote{While the $\k$ dependence of  $n$-vector 
couplings is fixed by dimensional analysis, their precise numerical coefficients may only be fixed by direct
calculations. }. 
Superficially, the existence of such an action may suggest that the $E_{7(7)}$ duality symmetry is 
consistent with the existence of UV divergences in this theory (\ref{cl}). One must however make sure that
the higher-order terms predicted by the covariant duality construction are the same as those obtained 
from direct calculations. Not being able to render them consistent ({\it e.g.} by adjusting the deformation 
source) would imply that some of the assumptions of the construction ({\it e.g.} that the tree-level duality 
transformations are unmodified) may need to be relaxed. 

   \item construction of a Born-Infeld type $\cN=8$ supergravity is not
     possible. If it is  possible to simultaneously prove that the classical
     $E_{7(7)}$ transformations do not receive modifications at the quantum
     level, then $E_{7(7)}$ would predict UV finiteness of $\NeqEight$
     supergravity in four dimensions.

\end{itemize}

Either outcome would expose more of the quantum properties of $\NeqEight$ supergravity 
and the consequence classical duality symmetries have on them. Similar scenarios exist in 
all theories exhibiting duality-invariant equations of motion. Along the way to $\NeqEight$ 
supergravity, consideration of such symmetries may lead to the identification of other supergravity 
theories with unexpectedly good ultraviolet properties or may point to a mechanism that makes 
duality symmetries consistent with the existence of counterterms/UV divergences.


\section*{Acknowledgments}
We are grateful to S.~Ketov and S.~Kuzenko and S.~Theisen for their help in understanding the issues of $\NeqTwo$ supersymmetry and self-duality and for providing us with important details of the work \cite{Kuzenko:2000tg}. We are grateful to P.~Aschieri, W.~Chemissany, T.~Ortin and A.~Tseytlin for the stimulating discussions.
We are grateful to A.~Linde for raising the issue of the predictive power of the two-coupling model (\ref{BIs}) with regard to the one coupling model in (\ref{cl}).
J.B.~acknowledges support from the Alexander-von- Humboldt foundation within the Feodor-Lynen program. J.J.M.C. and R.K. gratefully acknowledge the Stanford Institute for Theoretical Physics and the NSF Grant No. 0756174 for  their support. The work of S.F.~has been supported by the ERC Advanced Grant no. 226455, ÒSupersymmetry, Quantum Gravity and Gauge FieldsÓ (SUPERFIELDS), and in part by DOE Grant DE-FG03-91ER40662. R.R.~also acknowledges the support of NSF under Grant No.~0855356.
\appendix

\section{Review of $\boldsymbol{{\cal N}=0}$ and $\boldsymbol{{\cal N}=1}$ duality}
\label{appA}

A manifestly ${\cal N}=1$ supersymmetric NGZ-type identity derived by Kuzenko and Theisen 
in \cite{Kuzenko:2000tg}, where it was called ``$\NeqOne$ self-duality equation,'' is
\be 
\int {\rm d}^6 z\, 
\Big( W^\a W_\a + M^\a M_\a \Big) ~=~
\int {\rm d}^6 {\bar z}\,
\Big( {\overline W}_\ad{\overline W}^\ad  +
{\overline M}_\ad {\overline M}^\ad \Big) ~.
\label{N=1dualeq}
\ee
where the chiral and antichiral $\cN=1$ superfield strengths are defined as 
\be
W_\a=-\frac{1}{4}{\overline D}^2 D_\a V\quad{\rm and}\quad {\overline
W}_\ad=\frac{1}{4}D^2{\overline D}_\ad V
\ee
in terms of a real unconstrained prepotential $V$. In analogy to \eqn{N=2vd} one
defines
\be
{\rm i} M_\a \equiv 2\, \frac{\d }{\d W^\a}\,
{S}[W , {\overline W}]
 \, , \qquad \quad
- {\rm i} \overline{M}^\ad \equiv 2\, \frac{\d }{\d \overline{W}_\ad}\,
{S}[W , {\overline W}]\ .
\label{N=1vd}
\ee

${\cal N}=1$ duality invariant models can be obtained by considering a general
action of the form 
\be 
S ~=~ \frac{1}{4}\int {\rm d}^6z \, W^2 +
\frac{1}{4}\int {\rm d}^6{\bar z} \,{\overline  W}^2 
+  \frac{1}{4}\, \int {\rm d}^8z \, W^2\,{\overline W}^2  \,
\cL \Big(\frac{1}{8} D^2\,W^2 , \frac{1}{8}{\overline D}^2\, {\overline W}^2 \Big)~,
\label{gendualaction}
\ee
where $\cL(u,\bar u )$ is a real analytic function 
of the complex variable $u \equiv \frac{1}{8}D^2\,W^2$ and 
its conjugate. With \footnote{Here, as well as in \cite{Kuzenko:2000uh}, the
$\cN=1$ functional superderivative is defined as $\frac{\delta
W^\b(z')}{\delta W^\a(z)}=-\frac{1}{4}\delta_\a^{\ \b}{\overline D}^2\,\delta^8(z-z')$.}
\be
\frac{\delta S}{\delta W^\a}=\frac{1}{2}W_\a\Big(1-\frac{1}{4}{\overline D}^2\Big[{\overline
W}^2\Gamma\Big]\Big)\quad{\rm where}\quad \Gamma=\cL+\frac{1}{8}D^2\Big[W^2\frac{\delta
\cL}{\delta u}\Big]
\ee
one finds with \eqn{N=1vd}
\be
{\rm i}\,M_\a ~=~ W_\a \Big(1-\frac{1}{4}{\overline D}^2\Big[{\overline
W}^2\Gamma\Big]\Big)\, 
\ee
or, equivalently, 
\be
{\rm i}\,M_\a ~=~ W_\a \, \Bigg\lbrace\;
1-\frac{1}{4}\,{\overline D}^2  \Big[ {\overline W}^2  \Big( \cL +\frac{1}{8} 
D^2 \Big( W^2 \frac{\pa \cL(u, \bar u)}{\pa u} \Big)  \Big)  \Big]
\;  \Bigg\rbrace~.
\label{N=1M}
\ee

Plugging \eqn{N=1M} into the NGZ constraint \eqn{N=1dualeq} leads to a
functional equation for $\Gamma$
\be
\int{\rm d}^8 z W^2 {\overline W}^2{\rm Im}[\Gamma-{\overline u} \Gamma^2]=0\ ,
\ee
where we have used that for any $\cN=1$ superfield $Y$
\be
W^2{\overline W}^2{\overline D}^2[{\overline W}^2\,Y]={W^2{\overline
W}^2\overline D}^2[{\overline W}^2]\,Y\,
\label{specialN=1}
\ee
because $W^3=W_\a W_\b W_\g=0$ for the two-component spinor $W_\a$. One can
rewrite the above constraint in terms of $\cL$ as
\be
\int{\rm d}^8 z W^2 {\overline W}^2{\rm
Im}\Big[\partial_u(u\cL)-\overline{u}(\partial_u(u\cL))^2\Big]=0\ .
\label{N=1constraint}
\ee
This partial differential equation has infinitely many solutions, parametrized
e.g. by the coefficients of the terms $(u\overline{u})^n$ with $n>2$ in the
expansion around $u=0$ (as well as the coefficient of $u\overline{u}^2$), as was
shown in \cite{Gibbons:1995cv} in the non-supersymmetric case.

The relation to the non-supersymmetric case discussed in \cite{CKR} is
straightforward: taking the integral over the fermionic superspace coordinates
and setting the gauginos and auxiliary fields to zero, one finds
\be
L=-\frac{1}{2}( {\bf u}+\overline{\bf u})+{\bf u}{\overline{\bf u}}\,\cL({\bf
u}\overline{\bf u}),\qquad{\bf u}  \,  \equiv  \,  \frac{1}{8} D^2 W^2 \big|_{\q = 0, D=0,\psi=0}  \, \Rightarrow  \,  
\frac{1}{4}F^2+\frac{i}{4}F{\tilde F}\equiv\o\ .
\ee
In the non-supersymmetric cases the Born-Infeld (BI) and Bossard-Nicolai (BN)
examples are reproduced by functions \cite{CKR}:
\bea
\cL_{\rm BI}  &=&
\frac{g^2 }{ 1 + 
\hf g^2(\o + \bar \o )
+ \sqrt{1 + g^2 (\o + \bar \o )  
+\frac{1}{4}g^4 (\o - \bar \o )^2 }} \\
&=& \frac{g^2}{2}
     -\frac{g^4}{4}(\o+\bar\o)+\frac{g^6}{8} ( (\o+\bar\o)^2+\o\bar\o
     )-\frac{g^8}{16}( (\o+\bar\o)^3 + 3 (\o^2\bar\o+\o\bar\o^2))
     +\dots
\eea
and
\bea
\cL_{\rm BN} &=&\frac{g^2}{2} 
     -\frac{g^4}{4}(\o+\bar\o)+\frac{g^6}{8} ( (\o+\bar\o)^2+2\o\bar\o )
     -\frac{g^8}{16}\bigl((\o+\bar\o)^3+7(\o^2\bar\o+\o\bar\o^2) \bigr)+\dots\ 
\nonumber
\eea
where the first deviation occurs at ${\cal O}(g^6)$.

With the above identifications, the same is true for the model with ${\cal N}=1$ supersymmetry. At ${\cal
O}(g^6)$ a deviation between the Born-Infeld-type $\cN=1$ model and the $\cN=1$
supersymmetrization of the Bossard-Nicolai model will occur:
\be
\cL_{\rm BN}|_{ {\cal O}(g^6)}-\cL_{\rm BI}|_{ {\cal O}(g^6)}  = \frac{g^6}{8}\o\bar\o\equiv
\frac{g^6}{8}D^2W^2\overline{D}^2\overline{W}^2\,.
\ee
Comparing now the ${\cal O}(\lambda^3)$ terms in \eqn{ExampleAAction}, which
corresponds to the $\cN=2$ supersymmetrization of the BN-initial deformation, with the
corresponding terms in the Born-Infeld action \eqn{n2BIsoln},
one finds again the difference
\be
\frac{\lambda^3}{32}{\cal H}^{(0)}{\overline{\cal
H}}^{(0)}=\frac{\lambda^3}{32}\SD \cW^2\SDbar\cWbar^2\,,
\ee
where we have set $a=-\frac{1}{16}$ in \eqn{ExampleAAction} in order to connect
to the BI result as discussed in subsection \ref{sec:I2}.

\section{$\boldsymbol{{\cal N}=2}$ SUSY, $\boldsymbol{U(1)}$ duality, and derivative requirements }
\label{appB}

Here we consider a generalization of the ${\cal N}=0$ and ${\cal N}=1$ actions
assuming---as in \cite{Kuzenko:2000tg}---dependence on the $\cN=2$ superfields $\cW^2$ and
${\overline\cW}^2$. While the free action reads 
\bea
\cS_{\rm free} =
{ 1 \over 8} \,  \int {\rm d}^{8} \cZ \, 
\cW^2 + { 1 \over 8} \int {\rm d}^{8} \cZbar \,{\cWbar}^2\, \, , 
\eea
it is obvious to try, whether an ansatz similar\footnote{The prefactor is chosen
to allow for straightforward comparison of the resulting differential equation
with the $\cN=1$ model in the last section. Of course, any
prefactor can be absorbed in the definition of $\cL$.} to the interaction part of
\eqn{gendualaction},
\bea
\cS_{\rm int} =
-{ 1 \over 4} \,  \int {\rm d}^{12} \cZ \, 
\cW^2\,{\cWbar}^2\,\cL(\cD^4 \cW^2, \SDbar \cWbar^2)
\label{N2action} 
\eea
suffices to produce a duality-invariant action. We will show in the following,
that this is not the case: one has to include new terms depending on spacetime derivatives in order to
satisfy conservation of the NGZ current~\eqn{N=2dualeq}. 

Using \eqn{N=2vd} in the above equation yields
\be
\frac{\delta \cS_{\rm int}}{\delta \cW}=-\frac{1}{2}\cW\SDbar(\cWbar^2\Gamma)\quad{\rm
where}\quad \Gamma=\cL+\SD\Big[\cW^2\frac{\delta
\cL}{\delta u}\Big]\ .
\ee
where here $u=\SD\cW^2$. In terms of $\Gamma$ one finds
\be
i\,\cM=\cW\Big(1-2\cW\SDbar[\cWbar^2\Gamma]\Big)\ .
\label{N=2MSol}
\ee
Plugging \eqn{N=2MSol} $\cM$ into the self-duality equation
(\ref{N=2dualeq}) yields 
\be
\int {\rm d}^{12} \cZ \, 
\cW^2\,{\cWbar}^2\, \rm{ Im } [\Gamma - \Gamma \SDbar (\cWbar^2\Gamma)]=0\,.
\ee
Rewriting the above equation in terms of $\cL$ does not give the same beautiful
result as in the $\cN=1$ situation: there is no $\cN=2$-analogue to
\eqn{specialN=1}. As the $\cN=2$ superfield $\cW$ is a scalar, terms containing
$\cW^3$ will not vanish. Thus the constraint analogue to \eqn{N=1constraint} reads 
\be
\int {\rm d}^{12} \cZ \, 
\cW^2\,{\cWbar}^2\, \rm{ Im }
\Big[\partial_u(u\cL) - \overline{u}(\partial_u(u\cL))^2+\Delta\cL\Big]=0\,.
\ee
The correctional term $\Delta \cL$ contains terms like $\SD[\SDbar\cWbar]$
which, after carrying out the derivatives, yields terms proportional to
$\partial_\mu \cWbar$ which do not cancel.
Thus, the ansatz \eqn{N2action} is not sufficient, instead one needs
\bea
\cS_{\rm int} =
{ 1 \over 2} \,  \int {\rm d}^{12} \cZ \, 
\cW^2\,{\cWbar}^2\,\cL(\cD^4 \cW^2, \SDbar \cWbar^2)
+\cO (\partial \cW, \partial \cWbar)\ .
\label{N2actioncorrect} 
\eea
Comparing with \eqn{Ketov}, one finds that those terms do indeed appear. Already at
${\cal O}(\lambda^2)$ there is a term $\frac{\lambda^2}{9}\cW^3\Box\cWbar^3$,
which would have vanished in a $\cN=1$ supersymmetric theory. At the next order,
${\cal O}(\lambda^3)$, one finds terms of the form $\cW^2 \cWbar^2
\SD[\cW^2\SDbar[\cWbar^2]]$, which are not covered by an ansatz of the form
$\cW^2\cWbar^2\cL(\SD\cW^2,\SDbar\cWbar^2)$.

\section{Conventions of $\boldsymbol{{\cal N}=2}$ superspace}
\label{conventionsSUSY}

Superderivatives are defined as
\be
{\cal D}_\a^i=\partial_\a^i+{\rm i}\bar\theta^{\ad
i}\partial_{\a\ad}\qquad{\rm and}\qquad
\overline{\cal D}_{\ad i}=-\partial_{\ad i}-{\rm
i}\theta^\a_i\partial_{\a\ad}
\ee
where $\a, \ad$ are usual $SU(2)$ spinor indices and latin indices are
super-indices in the range from $\{1,\ldots 4\}$.  Anticommutation relations read
\be
\{ {\cal D} ^i_\a,\overline{\cal D}_{\ad i}\}=-2\;{\rm
i}\;\delta_j^i\;\partial_{\a\ad}\,.
\ee
Derivatives can be combined into 
\be
{\cal D}^{ij}={\cal D}^{\a i}{\cal D}^j_\a\qquad{\rm and}\qquad\overline{\cal
D}^{ij}=\overline{\cal D}^i_\ad\overline{\cal D}^{\ad j}
\ee
and finally
\be
{\cal D}^4=\frac{1}{48}{\cal D}^{ij}{\cal
D}_{ij}\qquad{\rm and}\qquad\overline{\cal D}^4=\frac{1}{48}\overline{\cal D}^{ij}\overline{\cal
D}_{ij}\ .
\ee
Chiral and antichiral superfields $\cW(x,\theta)$ and $\cWbar(x,\bar\theta)$ are defined as 
\be
\overline{\cal D}_{\ad i}\cW=0\qquad{\rm and}\qquad{\cal D}^i_\a\cWbar=0\,.
\ee
For a full superfield ${\cal V}(x,\theta,\bar\theta)$,
\be
\int{\rm d}^8 \cZ \,\SDbar{\cal V}=\int{\rm d}^8 \cZ \,\cW=\int{\rm d}^{12} \cZ \,{\cal V}
\ee
Correspondingly, the functional derivative for chiral and antichiral superfields are defined via 
\be
\frac{\delta\cW(\cZ)}{\delta\cW(\cZ')}=\SDbar\delta^{12}(\cZ-\cZ')\qquad{\rm
and}\qquad\frac{\delta\cWbar(\cZ)}{\delta\cWbar(\cZ')}=\SD\delta^{12}(\cZ-\cZ').
\ee
Because of anticommutativity, powers higher than four in the superderivatives
vanish (here we write the chiral part only, the
antichiral is completely equivalent),
\be
{\cal D}^n=0\quad\forall n>4\,,
\ee
which leads to 
\be
\SD(\SD(x))=0\qquad{\rm and\ thus}\qquad\SD(X\SD(Y))=\SD(X)\SD(Y)
\ee
for arbitrary $X$ and $Y$. Besides linearity and scaling 
\be
\SD(X+Y)=\SD(X)+\SD(Y)\quad{\rm and}\quad\SD(cX)=c\SD(X)\quad\forall\,{\rm scalar}
\,c
\ee
the chain rule does not apply trivially due to the product structure of $\SD$.

For the space-time d'Alambert operator appearing in the examples in subsections \ref{I2} and \ref{I4}, the following
commutation relation holds:
\be
\Box\SD(X)=\SD\Box(X)\,.
\ee
%



\begin{thebibliography}{10}

\bibitem{Ferrara:1976iq} 
  S.~Ferrara, J.~Scherk and B.~Zumino,
  ``Algebraic Properties of Extended Supergravity Theories,''
  Nucl.\ Phys.\ B {\bf 121}, 393 (1977).


  
\bibitem{Cremmer:1978km}
  E.~Cremmer, B.~Julia, J.~Scherk,
  ``Supergravity Theory in Eleven-Dimensions,''
  Phys.\ Lett.\  {\bf B76}, 409-412 (1978).

  E.~Cremmer and B.~Julia, 
``The SO(8) Supergravity,"
  Nucl.\ Phys.\  B {\bf 159}, 141 (1979).

  B.~de Wit and H.~Nicolai, 
 ``${\cal N}=8$ Supergravity,"
  Nucl.\ Phys.\  B {\bf 208}, 323 (1982).

  B.~de Wit, 
``Properties Of SO(8) Extended Supergravity,"
  Nucl.\ Phys.\  B {\bf 158}, 189 (1979).

  B.~de Wit and D.~Z.~Freedman, 
``On SO(8) Extended Supergravity,"
  Nucl.\ Phys.\  B {\bf 130}, 105 (1977).
  
  
\bibitem{N8Calculation}
  Z.~Bern, J.~J.~Carrasco, L.~J.~Dixon, H.~Johansson, D.~A.~Kosower and R.~Roiban, 
  ``Three-Loop Superfiniteness of ${\cal{N}}=8$ Supergravity,"
  Phys.\ Rev.\ Lett.\  {\bf 98}, 161303 (2007)
  [arXiv:hep-th/0702112].

  Z.~Bern, J.~J.~M.~Carrasco, L.~J.~Dixon, H.~Johansson, R.~Roiban,
  ``Manifest Ultraviolet Behavior for the Three-Loop Four-Point Amplitude of N=8 Supergravity,''
  Phys.\ Rev.\  {\bf D78}, 105019 (2008).
  [arXiv:0808.4112 [hep-th]].

  Z.~Bern, J.~J.~M.~Carrasco, H.~Johansson,
  ``Perturbative Quantum Gravity as a Double Copy of Gauge Theory,''
  Phys.\ Rev.\ Lett.\  {\bf 105}, 061602 (2010).
  [arXiv:1004.0476 [hep-th]].
 
\bibitem{Brodel:2009hu}
  J.~Broedel and L.~J.~Dixon,
  ``R$^4$ Counterterm andE \textbf{7} (7) Symmetry in Maximal Supergravity,''
  JHEP {\bf 1005} (2010) 003
  [arXiv:0911.5704 [hep-th]].
  
  H.~Elvang and M.~Kiermaier,
  ``Stringy KLT relations, global symmetries, and $E_{7(7)}$ violation,''
  JHEP {\bf 1010}, 108 (2010)
  [arXiv:1007.4813 [hep-th]].

\bibitem{Beisert:2010jx}
  N.~Beisert, H.~Elvang, D.~Z.~Freedman, M.~Kiermaier, A.~Morales and S.~Stieberger,
  ``E7(7) Constraints on Counterterms in ${\mathcal{N}}\!=8$ Supergravity,''
  Phys.\ Lett.\ B {\bf 694} (2010) 265
  [arXiv:1009.1643 [hep-th]].

\bibitem{Gaillard:1981rj}
  M.~K.~Gaillard and B.~Zumino, 
  ``Duality Rotations For Interacting Fields,"
  Nucl.\ Phys.\  B {\bf 193}, 221 (1981).

  P.~Aschieri, S.~Ferrara and B.~Zumino, 
  ``Duality Rotations in Nonlinear Electrodynamics and in Extended
  Supergravity,"
  Riv.\ Nuovo Cim.\  {\bf 31}, 625 (2008)
  [arXiv:0807.4039 [hep-th]].

\bibitem{Kallosh:2011dp}
  R.~Kallosh, 
  ``$E_{7(7)}$ Symmetry and Finiteness of ${\cal N}=8$ Supergravity,"
  arXiv:1103.4115 [hep-th].

  R.~Kallosh, 
  ``${\cal N}=8$ Counterterms and $E_{7(7)}$ Current Conservation,"
  JHEP {\bf 1106}, 073 (2011)
  [arXiv:1104.5480 [hep-th]].
  
\bibitem{Bianchi:2008pu}
  M.~Bianchi, H.~Elvang and D.~Z.~Freedman,
  ``Generating Tree Amplitudes in N=4 SYM and N = 8 SG,''
  JHEP {\bf 0809}, 063 (2008)
  [arXiv:0805.0757 [hep-th]].

  N.~Arkani-Hamed, F.~Cachazo and J.~Kaplan,
  ``What is the Simplest Quantum Field Theory?,''
  JHEP {\bf 1009}, 016 (2010)
  [arXiv:0808.1446 [hep-th]].

  R.~Kallosh and T.~Kugo,
 ``The footprint of E7 in amplitudes of N=8 supergravity,''
  JHEP {\bf 0901}, 072 (2009)
  [arXiv:0811.3414 [hep-th]].

\bibitem{Tseytlin:1991wr} 
  A.~A.~Tseytlin,
  ``Duality and dilaton,''
  Mod.\ Phys.\ Lett.\ A {\bf 6}, 1721 (1991).

\bibitem{Buscher:1987qj}
  T.~H.~Buscher,
  ``Path Integral Derivation of Quantum Duality in Nonlinear Sigma Models,''
  Phys.\ Lett.\ B {\bf 201} (1988) 466.

\bibitem{BN}
  G.~Bossard, H.~Nicolai,
 ``Counterterms vs. Dualities,''
  JHEP {\bf 1108}, 074 (2011).
  [arXiv:1105.1273 [hep-th]].

\bibitem{CKR}
  J.~J.~M.~Carrasco, R.~Kallosh, R.~Roiban,
  ``Covariant procedures for perturbative non-linear deformation of duality-invariant theories,''
  [arXiv:1108.4390 [hep-th]].
  
\bibitem{CKO} 
  W.~Chemissany, R.~Kallosh and T.~Ortin,
  ``Born-Infeld with Higher Derivatives,''
  arXiv:1112.0332 [hep-th].
   
\bibitem{Cecotti:1986gb}
  S.~Cecotti and S.~Ferrara, 
  ``Supersymmetric Born-Infeld Lagrangians,"
  Phys.\ Lett.\  B {\bf 187}, 335 (1987).
  
\bibitem{Bagger:1996wp} 
  J.~Bagger and A.~Galperin,
  ``A New Goldstone multiplet for partially broken supersymmetry,''
  Phys.\ Rev.\ D {\bf 55}, 1091 (1997)
  [hep-th/9608177].

  M.~Rocek and A.~A.~Tseytlin,
 ``Partial breaking of global D = 4 supersymmetry, constrained superfields, and three-brane actions,''
  Phys.\ Rev.\ D {\bf 59}, 106001 (1999)
  [hep-th/9811232].

\bibitem{Ket1}  
S.V.~Ketov,
``A Manifestly N=2 supersymmetric Born-Infeld action,'' 
Mod.Phys.Lett.A14:501-510 (1999)
  
\bibitem{Ket2}
  S.V.~Ketov, 
  ``Born-Infeld-Goldstone superfield 
  actions for gauge-fixed D-5 and D-3  branes in 6d,"
  Nucl.\ Phys.\  {\bf B553}, 250 (1999) 
  [hep-th/9812051].

\bibitem{Kuzenko:2000tg}
  S.~M.~Kuzenko and S.~Theisen, 
  ``Supersymmetric duality rotations,"
  JHEP {\bf 0003}, 034 (2000)
  [arXiv:hep-th/0001068].
 
\bibitem{Kuzenko:2000uh}
  S.~M.~Kuzenko and S.~Theisen, 
  ``Nonlinear selfduality and supersymmetry,"
  Fortsch.\ Phys.\  {\bf 49}, 273 (2001)
  [arXiv:hep-th/0007231].
  
\bibitem{Bellucci:2001hd}
  S.~Bellucci, E.~Ivanov, S.~Krivonos,
 ``Towards the complete {\cal N}=2 superfield Born-Infeld action with partially broken N=4 supersymmetry,''
  Phys.\ Rev.\  {\bf D64}, 025014 (2001).
  [hep-th/0101195].

\bibitem{Ketov:2001dq}
  S.~V.~Ketov, 
  ``Many faces of Born-Infeld theory,"
  arXiv:hep-th/0108189.
  
\bibitem{Andrianopoli:1996ve} 
  L.~Andrianopoli, R.~D'Auria and S.~Ferrara,
  ``U duality and central charges in various dimensions revisited,''
  Int.\ J.\ Mod.\ Phys.\ A {\bf 13}, 431 (1998)
  [hep-th/9612105].

\bibitem{Shmakova:1999ai}
M.~Shmakova,
  ``One loop corrections to the D3-brane action,''
  Phys.\ Rev.\ D {\bf 62}, 104009 (2000)
  [hep-th/9906239].
  
\bibitem{DeGiovanni:1999hr} 
  A.~De Giovanni, A.~Santambrogio and D.~Zanon,
  ``$\alpha'{}^4$ corrections to the N=2 supersymmetric Born-Infeld action,''
  Phys.\ Lett.\ B {\bf 472}, 94 (2000)
  [Erratum-ibid.\ B {\bf 478}, 457 (2000)]
  [hep-th/9907214].
    
\bibitem{us} J.~Broedel, J.~J.~M.~Carrasco, S.~Ferrara, R.~Kallosh, R.~Roiban, work in progress

\bibitem{Kallosh:1980fi}
  R.~E.~Kallosh,
  ``Counterterms in extended supergravities,''
  Phys.\ Lett.\  B {\bf 99} (1981) 122;

  P.~S.~Howe and U.~Lindstrom,
  ``Higher Order Invariants In Extended Supergravity,''
  Nucl.\ Phys.\  B {\bf 181}, 487 (1981).

\bibitem{Gibbons:1995cv}
  G.~W.~Gibbons and D.~A.~Rasheed, 
  ``Electric - magnetic duality rotations in nonlinear electrodynamics,"
  Nucl.\ Phys.\  B {\bf 454}, 185 (1995)
  [arXiv:hep-th/9506035].

  M.~K.~Gaillard and B.~Zumino, 
  ``Nonlinear electromagnetic self-duality and Legendre transformations,"
  arXiv:hep-th/9712103.

\end{thebibliography}
\end{document}